\documentclass[12pt]{article}

\usepackage{latexsym}
\usepackage{amssymb,amsfonts,amsmath}
\usepackage{graphicx} 
\usepackage{indentfirst}
\usepackage{bbm}
\usepackage{amssymb}
\usepackage{verbatim}
\usepackage{amsmath, amsthm,amssymb}
\usepackage{mathrsfs}
\usepackage{hyperref}
\usepackage{amsfonts}
\usepackage{dsfont}
\usepackage{cite}
\usepackage{xcolor}
\usepackage[multiple]{footmisc}
\usepackage{slashed}

\parskip=\medskipamount

\usepackage[mathscr]{euscript}

\newcommand{\de}{{\nabla}}

\numberwithin{equation}{section}

\newcommand {\cC}{{\cal C}}
\newcommand {\cD}{{\cal D}}

\newcommand {\cF}{{\cal F}}

\newcommand {\cL}{{\cal L}}

\newcommand {\cN}{{\cal N}}

\def\a{\alpha}
\def\b{\beta}

\def\d{\delta}

\def\g{\gamma}

\def\l{\lambda}

\def\o{\omega}

\def\r{\rho}
\def\s{\sigma}

\def\z{\zeta}

\def\L{\Lambda}

\def\S{\Sigma}

\def\ri{{\rm i}}
\def\re{{\rm e}}

\newcommand{\gd}{{\dot\g}}
\newcommand{\dd}{{\dot\d}}
\newcommand{\ad}{{\dot{\alpha}}}
\newcommand{\bd}{{\dot{\beta}}}

\newcommand{\dbeta}{{\dot{\beta}}}

\newcommand{\ve}{\varepsilon}

\newcommand{\pa}{\partial}
\newcommand{\hf}{\frac12}

\newcommand{\psib}{\bar{\psi}}

\newcommand{\be}{\begin{equation}}
\newcommand{\ee}{\end{equation}}
\newcommand{\bea}{\begin{eqnarray}}
\newcommand{\eea}{\end{eqnarray}}
\newcommand{\non}{\nonumber}
\newcommand{\ba}{\begin{array}}
\newcommand{\ea}{\end{array}}

\def\double #1{#1{\hbox{\kern-2pt $#1$}}}

\newcommand{\ts}{{\tilde{\s}}}

\newcommand{\bsubeq}{\begin{subequations}}
\newcommand{\esubeq}{\end{subequations}}

\newcommand{\bphi}{{\bar\phi}}
\newcommand{\bpsi}{{\bar\psi}}

\newcommand{\tsigma}{\tilde{\sigma}}

\def\tr{{\rm tr}}

\expandafter\ifx\csname url\endcsname\relax
  \def\url#1{\texttt{#1}}\fi
\expandafter\ifx\csname urlprefix\endcsname\relax\fi
\providecommand{\eprint}[2][]{\url{#2}}

\begin{document}
\begin{titlepage}
\begin{flushright}
March, 2022 (latest revision September 2024)
\end{flushright}
\vspace{5mm}

\begin{center}
{\Large \bf 
Hyper-Dilaton Weyl Multiplet\\ 
of 4D, ${\mathcal{N}}=2$ Conformal Supergravity
}
\end{center}

\begin{center}

{\bf
Gregory Gold,
Saurish Khandelwal,
William Kitchin,\\
Gabriele Tartaglino-Mazzucchelli
} \\
\vspace{5mm}

\footnotesize{
{\it 
School of Mathematics and Physics, University of Queensland,
\\
 St Lucia, Brisbane, Queensland 4072, Australia}
}
\vspace{2mm}
~\\
\texttt{g.gold@uq.edu.au; s.khandelwal@uq.edu.au; w.kitchin@uq.net.au; g.tartaglino-mazzucchelli@uq.edu.au}\\
\vspace{2mm}

\end{center}

\begin{abstract}
\baselineskip=14pt
We define a new dilaton Weyl multiplet of ${\mathcal{N}}=2$ conformal supergravity in four dimensions. This is constructed by reinterpreting the equations of motion of an on-shell hypermultiplet as constraints that render some of the fields of the standard Weyl multiplet composite. The independent bosonic components include four scalar fields and a triplet of gauge two-forms. The resulting, so-called, hyper-dilaton Weyl multiplet defines a $24+24$ off-shell representation of the local ${\mathcal{N}}=2$ superconformal algebra. By coupling the hyper-dilaton Weyl multiplet to an off-shell vector multiplet compensator, we obtain one of the two minimal $32+32$ off-shell multiplets of ${\mathcal{N}}=2$ Poincar\'e supergravity constructed by M\"uller in 1986. On-shell, this contains the minimal ${\mathcal{N}}=2$ Poincar\'e supergravity multiplet together with a hypermultiplet where one of its physical scalars plays the role of a dilaton, while its three other scalars are dualised to a triplet of real gauge two-forms. Interestingly, a $BF$-coupling induces a scalar potential for the dilaton without a standard gauging.
\end{abstract}
\vspace{5mm}

\vfill
\end{titlepage}

\newpage
\renewcommand{\thefootnote}{\arabic{footnote}}
\setcounter{footnote}{0}

\tableofcontents{}
\vspace{1cm}
\bigskip\hrule


\allowdisplaybreaks

\section{Introduction}

Conformal supergravity has played an important role in several research 
avenues in the last five decades --- we refer the reader to a few books
for reviews and a more detailed list 
of references 
\cite{freedman,Lauria:2020rhc,SUPERSPACE,Buchbinder-Kuzenko}. 
The main aim of our paper is to revise some of the ingredients of the 
superconformal tensor calculus for matter-coupled Poincar\'e supergravity
focusing on the four-dimensional (4D), $\cN=2$ case, and more generally on theories with
eight real supercharges.
For instance, we will show how to define a new off-shell $24+24$ 
Weyl multiplet of $\cN=2$ conformal supergravity.

After the seminal papers on 4D,  $\cN=1$ supergravity
\cite{Kaku:1977pa,Ferrara:1977ij,Kaku:1978nz,Kaku:1978ea},
the superconformal tensor calculus
for the 4D,  $\cN=2$ case was first constructed in the 80s in
\cite{deWit:1979dzm,deWit:1980gt,deWit:1980lyi,deWit:1983xhu,deWit:1984rvr}
and also extended to the cases of 6D,  $\cN=(1,0)$ in \cite{Bergshoeff:1985mz};
5D,  $\cN=1$ in \cite{Kugo:2000hn,Fujita:2001kv,Kugo:2002vc,Bergshoeff:2001hc,Bergshoeff:2002qk,Bergshoeff:2004kh};
and more recently to three space-time dimensions in 
\cite{Butter:2013goa,Butter:2013rba}.
Similar to superspace approaches
(see \cite{SUPERSPACE,Buchbinder-Kuzenko} for introductory reviews
and, e.\,g., 
\cite{Howe1,Howe:1981gz,Galperin:1984av,Galperin:1987ek,Galperin:1987em,Galperin:2001seg,Kuzenko:2008ep,Kuzenko:2009zu,Butter_superspace:2011sr,Butter_reduction:2012xg,Butter:2014gha,Butter:2014xua,Butter:2015nza}
and references therein, 
for the 4D,  $\cN=2$ case)
a main advantage of the superconformal tensor calculus 
is to provide an off-shell
description of general supergravity-matter couplings.
This allows one to formulate general 
supergravity-matter couplings where supersymmetry is
engineered in a completely model independent way. 
The approach has been very successful in helping to decipher
many of the intricate geometrical structures associated to 
(two-derivatives) sigma-models in supergravity-matter 
systems with eight real supercharges, see e.\,g.,
\cite{deWit:1984wbb,Cremmer:1984hj,deWit:1999fp,deWit:2001brd,deWit:2001bk,Lauria:2020rhc}. 
The off-shell nature of the formalism has been a central 
ingredient in its employment to the study of supersymmetric 
localisation and supersymmetric quantum field theories 
on curved space-times --- see \cite{Pestun:2016zxk} 
for a recent extensive review.
Moreover, off-shell supersymmetry has also been a crucial 
ingredient when using the superconformal tensor calculus
to construct higher-derivative supergravity invariants
\cite{BSS1,LopesCardoso:1998tkj,Mohaupt:2000mj,Hanaki:2006pj,CVanP,Bergshoeff:2012ax,Butter:2013rba,Butter:2013lta,Kuzenko:2013vha,OP131,OP132,OzkanThesis,Butter:2014xxa,Kuzenko:2015jxa,BKNT16,Butter:2016mtk,BNT-M17,NOPT-M17,Butter:2018wss,Butter:2019edc,Hegde:2019ioy,Mishra:2020jlc}.
These play an important role, e.\,g., in the study 
of black-hole entropy in next to leading order 
AdS/CFT
--- see the recent works
\cite{Bobev:2020egg,Bobev:2021oku,Bobev:2021qxx} 
and references therein.

Within the superconformal tensor calculus, general 
supergravity-matter couplings are engineered by a few 
ingredients. 
First of all, one needs a conformal supergravity multiplet
--- named the \emph{Weyl} multiplet ---
which forms an off-shell 
representation of the local superconformal algebra
and contains the vielbein as one of its independent fields.
This multiplet defines the geometry 
(soft algebra) associated with the gauging of the superconformal
space-time symmetry.
Next, one identifies off-shell matter multiplets with 
local superconformal transformation rules in a Weyl multiplet
background. These two ingredients provide the kinematic 
data of a specific supergravity-matter system. 
Finally, one engineers locally superconformal invariant action
principles
constructed out of these multiplets to obtained well-defined
supergravity theories.\footnote{These tasks can be
simplified  by manifestly gauging the superconformal
algebra in superspace through the so-called 
\emph{conformal superspace}.
Conformal superspace was first introduced for 4D, $\cN=1,2$ supergravity in \cite{Butter:2009cp,Butter_superspace:2011sr}
(see also the seminal work  \cite{Kugo:1983mv})
and it was then developed for 
3D, $\cN$-extended supergravity \cite{Butter:2013goa},
5D, $\cN=1$ supergravity \cite{Butter:2014xxa},
and recently 6D, $\cN=(1,0)$ supergravity \cite{BKNT16,BNT-M17}.}

Assuming the matter multiplets contain
enough ``compensating'' degrees of freedom, one can  
suitably gauge fix part of the superconformal group, specifically dilatations, 
special conformal transformations, 
$S$-supersymmetry, and  $R$-symmetry, 
to obtain
supergravity models where only the 
super-Poincar\'e symmetry survives and is gauged.
For instance, pure 4D,  $\cN = 2$ Poincar\'e supergravity
can arise by the coupling of the 
\emph{standard Weyl multiplet}
\cite{deWit:1979dzm,deWit:1980gt,deWit:1980lyi,deWit:1983xhu,deWit:1984rvr}
to two compensating multiplets.
 There is significant freedom in doing so. 
 Typically, one uses 
 a vector multiplet and a hypermultiplet
 (e.\,g.,~a linear, non-linear, or hypermultiplet
 with or without a central charge)  as compensators
 --- see \cite{Lauria:2020rhc} for a recent review.
Note that, in this endeavour, for forty years 
the first step has predominantly been the same 
(standard Weyl multiplet), 
while most of the freedom that has been used concerned
the matter (compensators) side of this story.
However, it is natural to ask whether
it is possible to use alternative Weyl multiplets
and if it is useful to do so. These are the types of questions
that led to our paper.

Answers to these questions are already known.
For instance, if one considers the superconformal tensor calculus
for 6D,  $\cN=(1,0)$ supergravity, it has been known since 1986 
\cite{Bergshoeff:1985mz}
that there can be more than one Weyl multiplet.
In 6D,  the existence of a variant \emph{dilaton Weyl} multiplet
engineered as a standard Weyl multiplet coupled to an on-shell 
tensor multiplet has been a key ingredient to obtain
6D,  $\cN=(1,0)$ Poincar\'e supergravity 
by using superconformal techniques.
Similar ideas were employed to construct a variant 
dilaton Weyl multiplet for 5D,  $\cN=1$ conformal supergravity
as the standard Weyl multiplet coupled to an on-shell vector 
multiplet \cite{Bergshoeff:2001hc}.
Among interesting applications of these variant Weyl multiplets,
it is worth mentioning that 
both in 5D  and 6D  the use of the dilaton Weyl 
multiplet allowed the construction of the component actions for 
the supersymmetric extensions of 
all curvature squared combinations 
--- Riemann-squared, Ricci-squared, and scalar-curvature-squared
---
in  Poincar\'e supergravities,
see
\cite{BSS1,CVanP,Bergshoeff:2012ax,OP131,OP132,OzkanThesis,Butter:2014xxa,NOPT-M17,Butter:2018wss}.

For the 4D,  $\cN=2$ case 
the existence of a variant representation
of the Weyl multiplet of conformal supergravity
was argued in \cite{Siegel:1995px} and
was explicitly constructed only recently 
in \cite{Butter:2017pbp}.
For reasons that will soon be clear, we will refer to
this multiplet as the 
\emph{vector-dilaton Weyl} multiplet.
Its construction closely mimics the 5D  case \cite{Bergshoeff:2001hc}.
More specifically, 
in \cite{Butter:2017pbp} 
the system described by a 4D,  $\cN=2$ 
on-shell vector multiplet in a standard Weyl multiplet background
was interpreted as a new $24+24$ multiplet
of conformal supergravity.
Using the equations of motion for the vector multiplet, 
the existing covariant matter 
fields of the standard Weyl multiplet, i.e., the real antisymmetric tensor, $T_{ab}{}^{ij}$,
the real scalar field, $D$, and the spinor field, $\chi^i$,
together with the ${\rm U}(1)_R$ symmetry connection,
were traded for fields of the on-shell vector multiplet
\cite{Butter:2017pbp}. 
The complex scalar field, $X$,
of the vector multiplet 
then becomes 
an independent physical field whose real part 
plays the role of a dilaton
in a Poincar\'e supergravity constructed in this framework.
This Poincar\'e supergravity was  
constructed by coupling the vector-dilaton Weyl multiplet
to a $8+8$ linear multiplet compensator
\cite{Butter:2017pbp,Mishra:2020jlc}.
Upon gauge fixing dilatation, special conformal transformations, 
$S$-supersymmetry, and ${\rm U}(1)_R\times {\rm SU}(2)_R$
$R$-symmetry (up to a residual ${\rm U}(1)_R$),
the resulting $32+32$ Poincar\'e supergravity multiplet
comprises the following set of fields
\begin{align}
\{{e_m}^a,{\psi_{m}}{}_i^\alpha,\bar{\psi}_{m}{}^i_\ad, a_m|a_m',\l_\alpha^i,\bar\l^\ad_i, C, t_{mn}|v_m, t_{mn}', M, {b_{a}}{}^{ij}\}
~.
\end{align}
Here, $\{{e_m}^a,{\psi_{m}}{}_i^\alpha,\bar{\psi}_{m}{}^i_\ad, a_m\}$
are the fields of the $\mathcal{N}=2$ on-shell supergravity multiplet,
respectively: 
the vielbein, 
the gravitino and its conjugate,
and the real graviphoton gauge vector field. 
The fields
$\{v_m, t_{mn}', M, {b_{a}}{}^{ij}\}$,
that are respectively a real vector,
a real antisymmetric gauge two-form,
a real scalar,
and a triplet of real vectors,
are all auxiliary fields.
The remaining fields,
$\{a_m^{'},\l_\alpha^i,\bar\l^\ad_i, C, t_{mn}\}$,
are physical and describe an on-shell vector multiplet
where the imaginary part of the complex scalar field $X$
has been traded for a dual antisymmetric real gauge two-form, $t_{mn}$.

The previous $32+32$ off-shell Poincar\'e supergravity
turned out to coincide
with the one engineered in 1986 by M\"uller in
\cite{Muller_vector:1986ku} --- 
we will refer to this as the 
\emph{vector-M\"uller}
supergravity. 
It is useful to compare M\"uller's supergravity 
to the well-known $40+40$ off-shell supergravity of 
\cite{Fradkin-Vasiliev-N2,deWit:1979xpv,deWit:1979dzm,Breitenlohner-Sohnius-N2}. 
One initial feature is that, from the point of view of off-shell 
supersymmetry, M\"uller's multiplet is irreducible while
the $40+40$ multiplet is not. In fact, the vector-M\"uller multiplet
arises from a $24+24$ conformal supergravity multiplet coupled to a 
single $8+8$ off-shell compensator, 
while the $40+40$ multiplet requires two $8+8$ compensating multiplets.
The on-shell theories are however different. 
The $40+40$ off-shell supergravity
leads to a dynamical system containing only
the irreducible on-shell $\cN=2$
supergravity while M\"uller's $32+32$ off-shell supergravity 
comes  with 
an extra physical on-shell ``dilaton'' vector multiplet.

Interestingly, in 1986 M\"uller constructed another minimal $32+32$ off-shell
$\cN=2$ Poincar\'e supergravity \cite{Muller_hyper:1986ts}.
We will refer to this as the 
\emph{hyper-dilaton Poincar\'e} supergravity.
The multiplet of \cite{Muller_hyper:1986ts}
comprises the following fields
\begin{align}
\{{e_m}^a,{\psi_{m}}_i^\alpha,\bar{\psi}_{m}{}^i_\ad, a_m|
C,{t_{mn}}{}^{ij},\rho_\alpha^i,\bar{\rho}^\ad_i|
X^{ij}, W_{ab}, b_a\}
~.
\label{fields-hyper-muller}
\end{align}
Besides the fields of minimal 
$\mathcal{N}=2$ on-shell supergravity,
$\{{e_m}^a,{\psi_{m}}{}_i^\alpha,\bar{\psi}_{m}{}^i_\ad, a_m\}$, 
the fields
$\{X^{ij}, W_{ab}, b_a\}$
are respectively a real SU(2) triplet of Lorentz scalars,
a real antisymmetric tensor,
and a real vector.
These three are auxiliary fields \cite{Muller_hyper:1986ts}.
The remaining fields,
$\{C,{t_{mn}}{}^{ij},\rho_\alpha^i,\bar{\r}^\ad_i\}$,
are physical and describe an on-shell hypermultiplet
where three of the four hypermultiplet's scalar fields
have been traded for an SU(2) triplet of 
dual antisymmetric real gauge two-forms, 
$t_{mn}{}^{ij}=t_{mn}{}^{ji}=-t_{nm}{}^{ij}$.
Precisely as for the vector-M\"uller supergravity,
even though off-shell the hyper-dilaton Poincar\'e multiplet  is irreducible with $32+32$
degrees of freedom, 
the on-shell theory contains an extra 
$4+4$ dilaton multiplet which,
in this case, is described by a variant version
of a hypermultiplet where $C$ plays the role of a dilaton field.
Considering the recent superconformal description of the vector-M\"uller
supergravity \cite{Butter:2017pbp,Mishra:2020jlc}, it is natural to ask if
and how one can engineer the hyper-dilaton Poincar\'e supergravity by 
using the superconformal tensor calculus. 
The main aim of our paper is to show how this can be done by 
using a $24+24$ variant representation 
of the $\cN=2$ conformal supergravity
multiplet that we will refer to as 
the \emph{hyper-dilaton Weyl} multiplet.

The definition of the hyper-dilaton Weyl multiplet is fairly simple.
In fact, it closely mimics the description of
vector-dilaton Weyl multiplets
with the crucial difference that one starts with an on-shell hypermultiplet
\cite{Fayet:1975yi,FS2}
in a standard Weyl multiplet background
\cite{deWit:1980gt,deWit:1980lyi,deWit:1983xhu,deWit:1984rvr,deWit:1999fp},  
rather than with an on-shell
vector multiplet.
The constraints that arise by requiring the
algebra of local superconformal transformations 
to close on the fields of the hypermultiplet
can then be interpreted as algebraic equations for some of the fields
of the standard Weyl multiplet. 
More precisely, in the hyper-dilaton Weyl multiplet,
the standard Weyl multiplet's matter fields 
$(\S^{\a i},\bar{\S}_{\ad i})$ and $D$, 
together with the 
${\rm SU}(2)_R$ symmetry connection $\phi_m{}^{ij}$
become composite fields. 
On the other hand,
the four bosonic $q^{i\underline{i}}$ and four fermionic 
$(\rho_\a^{\underline{i}},\bar{\rho}^\ad_{\underline{i}})$ fields of the hypermultiplet,
together with an emerging triplet of real gauge two-forms 
$b_{mn}{}^{\underline{i}\underline{j}}
=b_{mn}{}^{\underline{j}\underline{i}}
=-b_{nm}{}^{\underline{i}\underline{j}}$,
are independent and not subject 
to any equations of motion.
By then coupling the $24+24$ hyper-dilaton Weyl multiplet
to a single $8+8$ off-shell vector multiplet compensator,
upon gauge fixing dilatation, special conformal transformations, 
$S$-supersymmetry, and the whole 
${\rm U}(1)_R\times {\rm SU}(2)_R$
$R$-symmetry,
one readily obtains the $32+32$ hyper-dilaton Poincar\'e supergravity
with the field content described in \eqref{fields-hyper-muller}.

To the best of our knowledge, despite their simplicity,
the reinterpretation that we advocate in this paper
for the on-shell hypermultiplet 
as the hyper-dilaton Weyl multiplet
and their connection to M\"uller's supergravity in a 
superconformal framework has never explicitly appeared 
before in the literature.
The advantage of our novel superconformal formulation
compared to the original work of 
M\"uller \cite{Muller_hyper:1986ts}
is the potentially
straightforward extension to more general matter couplings.
As a simple example, 
in our paper we show this by extending M\"uller's Poincar\'e
supergravity action by including a new invariant that leads
to a non-trivial potential for the dilaton field.
Intriguingly,
such a scalar potential is generated without 
a standard gauging which, 
in an $\cN=2$ standard Weyl multiplet setting, 
is associated to integrating out 
an independent auxiliary field
given by the ${\rm SU}(2)_R$ gauge connection.

This paper is organised as follows.
In section \ref{section-2} we first review the definition of 
the standard Weyl multiplet of  off-shell $\cN=2$
conformal supergravity and, by using our notation, 
present results we need for the rest of the paper.
We then describe the structure of an on-shell hypermultiplet
in a standard Weyl multiplet background and explain
how such a system can be reinterpreted as a variant
hyper-dilaton Weyl multiplet of off-shell $\cN=2$ 
conformal supergravity.
Section \ref{section-3} is devoted to first prove
how the off-shell Poincar\'e supergravity theory
constructed by M\"uller in \cite{Muller_hyper:1986ts}
can be engineered as the hyper-dilaton Weyl multiplet
coupled to an off-shell vector multiplet 
conformal compensator.
We then extend the results of \cite{Muller_hyper:1986ts}
by adding a new $BF$-coupling which induces 
a scalar potential for the dilaton without a standard 
$R$-symmetry gauging.
Section \ref{section-4} includes a final discussion
and an outline of some future directions
based on the results of our paper.


\section{The hyper-dilaton Weyl multiplet}
\label{section-2}

The aim of this section is to construct the $24+24$
hyper-dilaton Weyl multiplet of off-shell 
$\cN=2$ conformal supergravity.
Subsection \ref{SWM} reviews well-known results about the 
standard Weyl multiplet and serves to introduce
the notation that we employ. 
Subsection \ref{hyper+HDWM} describes the on-shell hypermultiplet 
in a standard Weyl multiplet background and the resulting
interpretation of this system
as an independent multiplet of conformal supergravity.

\subsection{The standard Weyl multiplet}
\label{SWM}

The standard Weyl multiplet  of 4D, $\cN=2$ conformal supergravity
is associated with the local off-shell gauging in space-time of the
superconformal group SU$(2,2|2)$ \cite{deWit:1979dzm},
see also 
\cite{deWit:1980gt,deWit:1980lyi,deWit:1983xhu,deWit:1984rvr}
and \cite{freedman,Lauria:2020rhc} for reviews.
The multiplet comprises $24+24$ physical components described by a set of independent gauge fields:
the vielbein $e_m{}^a$ and a dilatation connection $b_m$;
the gravitino $(\psi_m{}^\alpha_i,\bar\psi_m{}_\ad^i)$, associated with the gauging of $Q$-supersymmetry;
a U(1)$_R$ gauge field $A_m$;
and SU(2)$_R$ gauge fields $\phi_m{}^{ij}=\phi_m{}^{ji}$.
The fields associated to the remaining generators of SU$(2,2|2)$,
specifically the Lorentz connections $\o_m{}^{cd}$, 
$S$-supersymmetry connection $(\phi_m{}_\alpha^i,\bar\phi_m{}^\ad_i)$
and the special conformal connection $\mathfrak{f}_m{}_a$, are composite fields.
The connections define the locally superconformal covariant derivatives\footnote{We refer the reader to 
\cite{Butter_reduction:2012xg} for the notations that we adopt in our paper, 
and for the algebra of generators of SU$(2,2|2)$.
For instance, note that we use two-component 
SL$(2,\mathbb{C})$ spinor indices
according to the notation of \cite{Buchbinder-Kuzenko}.
We closely adhere to the conventions of
\cite{Butter_reduction:2012xg} with the exception of the overall signs in the definition of the connections, 
${\omega}_m{}^{cd}$,
${b}_m$,
$A_m$,
$\phi_m{}^{kl}$,
$({\phi}_m{}_{\a}^{ i},\bar{{\phi}}_m{}^{\ad}_{i})$,
and 
${\mathfrak f}_m{}_{a}$, 
as well as the overall signs for all the superconformal curvatures.}
\bea
{\nabla}_a
=e_a{}^m{\nabla}_m&=& 
e_a{}^m\Big(\pa_m
- \frac{1}{2} \psi_m{}^\a_i Q_{\a}^i 
- \frac{1}{2} \bar{\psi}_m{}_\ad^i \bar{Q}^{\ad}_i 
- \frac{1}{2}  {\omega}_m{}^{cd} M_{cd}
-\ri   A_m Y
-  \phi_m{}^{kl} J_{kl}
\non\\
&&~~~~~~
-  {b}_m \mathbb D
- \frac{1}{2}  {\phi}_m{}_{\a}^{ i} S^{\a}_{ i}
- \frac{1}{2}  \bar{{\phi}}_m{}^{\ad}_{i} \bar{S}_{\ad}^{ i}
-  {{\mathfrak f}}_m{}_{c} K^c
\Big)
~.
\eea
Together with the independent gauge connections, the standard Weyl 
multiplet includes a set of covariant matter fields 
that are necessary to close the local superconformal 
algebra off-shell: 
an anti-symmetric real tensor $W_{ab}=W^+_{ab}+W^-_{ab}$, 
which decomposes
into its imaginary-(anti-)self-dual components $W^{\pm}_{ab}$; 
a real scalar field $D$;
and fermions $(\Sigma^{\a i},\bar{\Sigma}_{\ad i})$.
The covariant derivatives satisfy the algebra 
\bea
{[}\de_a,\de_b{]}
&=&
-{R}(P)_{ab}{}^c\de_c
-{R}(Q)_{ab}{}^\a_i Q_\a^i
-{R}(\bar{Q})_{ab}{}_\ad^i \bar{Q}^\ad_i
- \frac{1}{2}  {R}(M)_{ab}{}^{cd} M_{cd}
-  {R}(\mathbb D)_{ab}\mathbb D
\non\\
&&
-\ri  R(Y)_{ab}Y
-  R(J)_{ab}{}^{kl} J_{kl}
- R(S)_{ab}{}_{\a}^{ i} S^{\a}_{ i}
- R(\bar{S})_{ab}{}^{\ad}_{i} \bar{S}_{\ad}^{ i}
- R(K)_{ab}{}_{c} K^c
~.
\eea
It is also useful to list the non-trivial 
conjugation properties
\bsubeq\bea
&(\psi_m{}^\a_i)^*=\bar{\psi}_m{}^{\ad i}
~,~~~
(\phi_m{}_\a^i)^*=\bar{\phi}_m{}_{\ad i}
~,~~~
(\phi_m{}^{ij})^*=\phi_m{}_{ij}
~,~~~
(\S^{\a i})^*=\bar{\S}^\ad_i
~,~~~~~~\\
&(R(Q)_{ab}{}^\a_i)^*=R(\bar{Q})_{ab}{}^{\ad i}
~,~~~
(R(S)_{ab}{}_\a^i)^*=R(\bar{S})_{ab}{}_{\ad i}
~,~~~
(R(J)_{ab}{}^{kl})^*=R(J)_{ab}{}_{kl}
~,~~~~~~
\eea\esubeq
while all the other fields and curvatures are real.

A set of conventional constraints express the superconformal curvatures in terms of connections and covariant matter fields and 
render the connections $\o_m{}^{cd}$, $(\phi_m{}_\alpha^i,\bar\phi_m{}^\ad_i)$,
and $\mathfrak{f}_m{}_a$ composite.
There is large freedom in the choice of conventional constraints, and, in fact,
different papers often make different choices.
In our paper we adopt the conventional constraints used in
\cite{Butter_superspace:2011sr}
adapted 
to our conventions. 
They are given by 
\bsubeq
\bea
R(P)_{ab}{}^c
&=&
0~,
\\
R(Q)_{ab}{}_j{\s}^b
&=&
-\frac{3}{4}\S_{j}\s_a
~,~~~~~~
R(\bar{Q})_{ab}{}^{j}{\tilde\s}^b
=
\frac{3}{4}\bar\S^{j}\tilde\s_a~,
\\
R(M)^c{}_{a}{}_{cb}
&=&
R(\mathbb D)_{ab}
+3\eta_{ab}D
-\eta^{cd}
W^-_{ac}W^+_{bd}
~.
\eea
\esubeq
Later in this section
we will present the expressions of the superconformal curvatures that
we need  while we refer the reader to
\cite{Butter_superspace:2011sr,Butter_reduction:2012xg}
for more detail and for the relation with the  
results of \cite{deWit:1979dzm}.

In presenting the multiplet we restrict
to all local superconformal transformations except local 
translations (covariant general coordinate transformations).
Such transformations are identified by $\d$ and defined by the following
operator
\bea
\delta
= 
\xi^\a_iQ_\a^i
+\bar{\xi}_\ad^i\bar{Q}^\ad_i
+\frac{1}{2}\lambda^{ab}M_{ab} 
+ \lambda^{ij}J_{ij} 
+ \lambda_{\mathbb{D}}\mathbb{D} 
+ \ri\lambda_Y Y 
+ \lambda_a K^a 
+ \eta_{\alpha}^i S^\alpha_i 
+ \bar{\eta}^{\dot{\alpha}}_i\bar{S}_{\dot{\alpha}}^i
~.
\eea
The local superconformal transformation of 
the fundamental fields of the standard Weyl multiplet
are  then given by
\bsubeq\label{transf-standard-Weyl}
\bea
\delta e_m{}^a 
&=& 
\ri\,\xi_i\sigma^a\bar{\psi}_m{}^i
+\ri\,\bar{\xi}^i \tilde\sigma^a\psi_m{}_{i}
- \lambda_{\mathbb{D}}{e}_m{}^a 
+ \lambda^{a}{}_{b}e_m{}^b
~,
\label{d-vielbein}
\\
\delta \psi_m{}^\alpha_i  
&=& 
\Big(2\partial_m \xi^\alpha_i  
+ {\omega}_m{}^{ab}(\xi_i\sigma_{ab})^\alpha 
+2\phi_m{}_{i}{}^{j}\xi^{\alpha}_{j} 
+ 2\ri A_m\xi^\alpha_i 
+ b_m\xi^\alpha_i
\Big)
-\frac{\ri}{2}(\bar{\xi}_{i}\tilde{\sigma}_m\s^{cd})^{\alpha}W^+_{cd}
\non\\
&&
-\frac{1}{2}\lambda^{ab}({\psi}_m{}_i\sigma_{ab})^\alpha 
- \lambda_{i}{}^{j} \,\psi_m{}^{\alpha}_{j}
- \ri\lambda_{Y}\,\psi_m{}^\alpha_i
- \frac{1}{2}\lambda_{\mathbb{D}}\,{\psi}_m{}^\alpha_i
+2\ri(\bar{\eta}_{i}\tilde{\sigma}_m)^{\alpha} 
~,
\label{d-gravitino}
\\
\delta 
\bar{\psi}_m{}_\ad^{i}  
&=&
\Big(
2\partial_m \bar{\xi}_\ad^{i}  
+{\omega}_m{}^{ab}(\bar{\xi}^{i} \tilde{\sigma}_{ab})_{\ad} 
- 2\phi_m{}^{i}{}_{j} \bar{\xi}_{\dot{\alpha}}^j 
- 2\ri A_m\bar{\xi}_\ad^{i} 
+ b_m \bar{\xi}_\ad^{i}
\Big)
+\frac{\ri}{2} (\xi^{i}{\sigma}_m\tilde{\s}^{cd})_{\dot{\alpha}} W^-_{cd}
\non
\\
&&
-\frac{1}{2}\lambda^{ab}({\bar{\psi}}_m{}^{i}\tilde{\sigma}_{ab})_{\ad}  
+ \lambda^{i}{}_{j} \,\bar{\psi}_m{}_{\dot{\alpha}}^j 
+ \ri\lambda_{Y} \,\bar{\psi}_m{}_\ad^{i}
- \frac{1}{2}\lambda^{\mathbb{D}}\,{\bar{\psi}}_m{}_\ad^{i}
+ 2\ri (\eta^{i}{\sigma}_m)_{\dot{\alpha}}
~,
\label{d-b-gravitino}
\\
 \delta  \phi_m{}^{ij} &=&
\Big(\partial_m\lambda^{ij} 
  -2\phi_m{}^{(i}{}_k \lambda^{j)k}
  \Big)
+\frac{3\ri}{2} \,\xi^{(i}{\sigma}_m\bar{\Sigma}^{j)}
 +\frac{3\ri}{2}\,\bar{\xi}^{(i}\tilde{\sigma}_m\Sigma^{j)}
 - \phi_m{}^{(i} \xi^{j)} 
+ \bar{\phi}_m{}^{(i}\bar{\xi}^{j)}
\non\\
&& 
 + 2 \psi_m{}^{(i}\eta^{j)} 
 - 2 \bar{\psi}_m{}^{(i} \bar{\eta}^{j)} 
~,
\label{d-SU2}
\\
\delta A_m 
&=& 
\partial_m\lambda_Y  
- \frac{3}{8}\,\xi_i{\sigma}_m\bar{\Sigma}^{i}
- \frac{3}{8}\,\bar{\xi}^i\tilde{\sigma}_m\Sigma_{i}
+ \frac{\ri}{2}\,\xi_i\phi_m{}^i
- \frac{\ri}{2}\,\bar{\xi}^i\bar{\phi}_m{}_i
- \frac{\ri}{2}\,\psi_m{}_i \eta^i
+\frac{\ri}{2}\,\bar{\psi}_m{}^i\bar{\eta}_i 
~,~~~~~~
\label{d-U1}
\\
\delta b_m 
&=& \partial_m \lambda_{\mathbb{D}} 
-\frac{3\ri}{4}\,\xi_i \sigma_m\bar{\Sigma}^{i} 
+ \frac{3\ri}{4}\,\bar{\xi}^i\tilde{\sigma}_m\Sigma_{i}
+\xi_i   \phi_m{}^i
+\bar{\xi}^i\bar{\phi}_m{}_i
- \psi_m{}_i\eta^i  
-\bar{\psi}_m{}^i\bar{\eta}_i 
- 2\lambda_m
~,
\label{d-dilatation}
\\
 \delta W_{ab} &=& 
-4\xi_{k}R(Q)_{ab}{}^k
-4\bar\xi^{k}R(\bar{Q})_{ab}{}_k
-2\lambda_{[a}{}^{c}W_{b]c}
  +\lambda_{\mathbb{D}} W_{ab}
  -2\ri \lambda_{Y} W^+_{ab}
  +2\ri \lambda_{Y} W^-_{ab}
  ~,~~~
  \label{d-W}
  \\
\delta  D
&=& 
-\ri \xi^{k} \s^a \de_{a}\bar{\Sigma}_k 
- \ri \bar{\xi}_k \tilde{\s}^a \de_a \Sigma^{k} 
+ 2 \l_{\mathbb{D}} D
~,
\label{d-D}
\\
\delta \Sigma^{\a i}
&=&      
\xi^{\a i }  D 
+ \frac{4\ri}{3} (\xi^{ i} \s^{ab})^\a R(Y)_{ab} 
+\frac{2}{3} (\xi_{j} \s^{ab})^\a R(J)_{ab}{}^{ j i} 
- \frac{\ri}{3} (\bar{\xi}^{i} \tilde{\sigma}^a\s^{cd})^\a
\nabla_{a}W^+_{cd}
\non\\ 
&& 
-\frac{1}{2} \lambda^{a b} (\Sigma^i \sigma_{ab})^{ \a} 
+\lambda^{i}{}_{ j}  \Sigma^{\a j} 
+\frac{3}{2}\lambda_{\mathbb{D}}
\Sigma^{\a i}
- \ri \lambda_{Y}\Sigma^{\a i}
+\frac{2}{3} (\eta^{i}\s^{cd})^{\alpha}W^+_{cd}
~,
\label{d-Sigma}
\\
\delta  \bar{\Sigma}_{\ad i}
&=& 
-\bar{\xi}_{\ad i} D 
+ \frac{4\ri}{3} (\bar{\xi}_{ i} \tilde{\s}^{ab})_{\ad} R(Y)_{ab} 
+ \frac{2}{3} (\bar{\xi}^{j} \tilde{\s}^{ab})_\ad R(J)_{ a b j i} 
- \frac{\ri}{3}(\xi_{ i} \s^a\tilde\s^{cd})_\ad \nabla_{a}W^-_{cd} 
\nonumber \\ 
&&  
- \frac{1}{2} \lambda^{a b}   (\bar{\Sigma}_i \tilde{\s}_{ab})_{\ad} 
- \lambda_{i}{}^{j}  \bar{\Sigma}_{\ad j}
+\frac{3}{2}\lambda_{\mathbb{D}} \bar{\Sigma}_{\ad i}
+ \ri\lambda_{Y} \bar{\Sigma}_{\ad i} 
+\frac{2}{3} (\bar{\eta}_{i}\tilde{\s}^{cd})_\ad W^-_{cd}
~,
\label{d-b-Sigma}
\eea
\esubeq
where
\bsubeq\label{nabla-on-W-and-Sigma}
\bea
\nabla_{a}W_{bc} &=& 
\cD_{a}W_{bc}
+2 \psi_{a}{}_{k}  R(Q)_{bc}{}^k
+2 \bar{\psi}_{a}{}^{k} R(\bar{Q})_{bc}{}_k 
~,\\
\de_a \Sigma^{\a i}
&=& 
\cD_a\Sigma^{\a i} 
-\frac{1}{2} \psi_{a}{}^{\a i}  D 
- \frac{2\ri}{3} (\psi_{a}{}^{i}\s^{cd})^\a  R(Y)_{cd}
- \frac{1}{3}(\psi_{a}{}_{j}\s^{cd})^\a R(J)_{cd}{}^{j i}  
\nonumber\\
&&
+ \frac{\ri}{6} (\bar{\psi}_{a}{}^{i}  \tilde{\s}^b\s^{cd})^\a \de_{b} W^+_{cd}
+ (\phi_{a}{}^{i}\s^{cd})^\a\, W^+_{cd}
~,
\\
\de_a \bar{\Sigma}_{\ad i}
&=& 
\cD_a\bar{\Sigma}_{\ad i} 
+\frac{1}{2} \bar\psi_{a}{}_{\ad i}  D 
- \frac{2\ri}{3}  (\bar\psi_{a}{}_{i}\tilde{\s}^{cd})_\ad  R(Y)_{cd}
- \frac{1}{3}(\bar\psi_{a}{}^{j}\tilde{\s}^{cd})_\ad R(J)_{cd}{}_{j i}  
\nonumber\\
&&
+ \frac{\ri}{6} (\psi_{a}{}_{i}\s^b\tilde{\s}^{cd})_\ad \de_{b} W^-_{cd} 
+  (\bar\phi_{a}{}_{i}\tilde{\s}^{cd})_\ad\, W^-_{cd}
~,
\eea
\esubeq
and the derivatives $\cD_a$ 
are\footnote{In many cases, including 
eqs.~\eqref{nabla-on-W-and-Sigma},
we will not explicitly write the expressions for
the $\cD_a$ derivatives acting 
on different fields. 
However, it is straightforward to obtain
the corresponding results knowing the 
chiral and dilatation weights, see Table 
\ref{chiral-dilatation-weights},
together with the Lorentz and SU(2)$_R$ 
representations of fields 
and by using the action of the generators 
$M_{ab}$ and $J_{ij}$
according to the notations of 
\cite{Butter_reduction:2012xg}.} 
\begin{align}
\cD_a 
=
e_a{}^m\cD_m
=
e_a{}^m\Big( \partial_m 
- \frac{1}{2}\omega_m{}^{cd}M_{cd} 
- \phi_m{}^{ij}J_{ij} 
- \ri A_m Y 
- b_m\mathbb{D}\Big)
~.
\end{align}
The composite Lorentz and $S$-supersymmetry connections are
respectively
\bea
\o_{abc}
&=&
\o(e)_{a}{}_{bc}
-2\eta_{a[b}b_{c]}
-\frac{\ri}{2}\big(
\psi_{a}{}_j\s_{[b}\bar{\psi}_{c]}{}^j
+\psi_{[b}{}_j\s_{c]}\bar{\psi}_a{}^j
+\psi_{[b}{}_j\s_{|a|}\bar{\psi}_{c]}{}^j
\big)
~,
\label{composite-Lorentz}
\eea
and
\bsubeq\label{composite-S-connection}
\bea 
\phi_{m}{}_\b^j
&=&
\frac{\ri}{4} \left(
\sigma^{bc} \sigma_m - \frac{1}{3} \sigma_m \tilde{\sigma}^{bc} 
\right)_{\b \dbeta}
\overline{\Psi}_{bc}{}^{\dbeta j}
+ \frac{1}{3} W_{mb}^- \psi^b{}_{\b}^j
- \frac{1}{3} W_{mb}^- (\sigma^{bc} \psi_{c}{}^j)_{\b}   
+ \frac{\ri}{4} (\sigma_{m} \bar\S^{j})_\b
~,
\\
\bphi_{m}{}^\bd_j 
&=& 
\frac{\ri}{4} \left(
\tsigma^{bc} \tsigma_m - \frac{1}{3} \tsigma_m \sigma^{bc} \right)^{\bd\beta}
\Psi_{bc}{}_{\beta j}
-\frac{1}{3} W_{mb}^+ \bar\psi^b{}^{\bd}_{j}
+ \frac{1}{3} W_{mb}^+ (\tsigma^{bc} \bar\psi_{c\, j})^{\bd}
- \frac{\ri}{4} (\tsigma_{m} \S_{j})^\bd
~.~~~~~~~~~~
\eea
\esubeq
The field $\o(e)_{a}{}_{bc}$ in \eqref{composite-Lorentz} 
is the usual torsion-free Lorentz connection 
given in terms of the anholonomy tensor 
${\cal{C}}_{mn}{}^a(e)$ as
\bea
\o(e)_{a}{}_{bc}
=
-\cC_{a}{}_{[bc]}(e)
+\hf\cC_{b}{}_{ca}(e)
~,~~~
{\cal{C}}_{mn}{}^a(e):=2\pa_{[m}e_{n]}{}^a
~,~
{\cal{C}}_{ab}{}^c(e)
:=
e_a{}^me_b{}^n{\cal{C}}_{mn}{}^c(e)
~,~~~~~
\eea
while the fields 
$\big(\Psi_{ab}{}^\g_k,\,{\overline{\Psi}}_{ab}{}^k_{\dot{\g}}\big)$
are the gravitini field strengths
\bea
\Psi_{ab}{}^\g_k 
= 
2{e_a}^m{e_b}^n\cD_{[m}\psi_{n]}{}^\g_k
~,~~~~~~
\overline{\Psi}_{ab}{}_\gd^k = 2{e_a}^m{e_b}^n\cD_{[m}\bar{\psi}_{n]}{}_\gd^k
~.
\eea
Note that $\big(R(Q)_{ab}{}^\g_k,\,R(\bar{Q})_{ab}{}_\gd^k\big)$
are the $Q$-supersymmetry curvatures and satisfy
\bsubeq
\bea
R(Q)_{ab}{}^\g_k
&=& \frac{1}{2} \Psi_{ab}{}^\g_k
- \ri (\bphi_{[a\, k} \,\tsigma_{b]})^{\g}
+\frac{\ri}{4}(\bpsi_{[a\,}{}_k \tsigma_{b]}\sigma^{cd})^{\g} W^+_{cd} 
~,
\\
R(\bar{Q})_{ab}{}_\gd^{k}
&=&
\frac{1}{2} \overline{\Psi}_{ab}{}_\gd^{k}
-\ri(\phi_{[a}{}^{k}\sigma_{b]})_{\gd}
-\frac{\ri}{4}(\psi_{[a}{}^{k}\sigma_{b]}\tsigma^{cd})_\gd W^-_{cd}
~,
\eea
\esubeq
while $R(Y)_{ab}$ and $R(J)_{ab}{}^{kl}$ are
\bsubeq
\bea
R(Y)_{ab} 
&=&
2 e_a{}^me_b{}^n  \partial_{[m} A_{n]}
- \frac{\ri}{2} \psi_{[a}{}_j\phi_{b]}{}^j
+ \frac{\ri}{2} \bpsi_{[a}{}^j\bphi_{b]}{}_j
+ \frac{3}{8} \psi_{[a}{}_j \sigma_{b]} \bar\S{}^j
+ \frac{3}{8} \bpsi_{[a}{}^j \tsigma_{b]} \S{}_j
~,~~~~~~
\\
R(J)_{ab}{}^{kl} 
&= &
2 e_a{}^m e_b{}^n\partial_{[m} \phi_{n]}{}^{kl}
-2\phi_{a}{}^{(k}{}_p
\phi_{b}{}^{ l) p}
+2\psi_{[a}{}^{(k}
\phi_{b]}{}^{ l)}
-2\bpsi_{[a}{}^{(k}
\bphi_{b]}{}^{l)}
\non\\
&&
- \frac{3\ri}{2} \psi_{[a}{}^{(k} \sigma_{b]}\bar\S^{l)}
- \frac{3\ri}{2} \bpsi_{[a}{}^{(k} \tsigma_{b]} \S^{l)}
~.
\eea
\esubeq
We are not going to present the expression for the composite special conformal connection $\mathfrak{f}_{m}{}_a$ and other superconformal
curvatures. We will however use $e_a{}^m\mathfrak{f}_{m}{}^a$ which 
is given by 
\bea
{\mathfrak{f}_a}^a 
&=&
- \frac{1}{12}R
+D 
- \frac{1}{24}\varepsilon^{mnpq}
({\bar{\psi}_m}^j\tilde{\sigma}_n\cD_p\psi_{q}{}_{j})
 +\frac{1}{24}\varepsilon^{mnpq}
 (\psi_{mj} \sigma_n \cD_p {\bar{\psi}_q}{}^j)
 \non\\
&&
-\frac{\ri}{8}\psi_{aj}\sigma^a \bar{\Sigma}^j 
+ \frac{\ri}{8}{\bar{\psi}_a}{}^j\tilde{\sigma}^a\Sigma_j
- \frac{1}{12}W^{ab+}({\bar{\psi}_a}{}^j\bar{\psi}_{bj})
+\frac{1}{12}W^{ab-}(\psi_{aj}{\psi_b}{}^j)
~,
\label{composite-f}
\eea
where $R=e_a{}^me_b{}^n R_{mn}{}^{ab}$ is the scalar curvature constructed from the Lorentz curvature
\bea
R_{mn}{}^{cd} 
=
2\pa_{[m}\o_{n]}{}^{cd}
-2\o_{[m}{}^{ce}\o_{n]}{}_e{}^{d}
~.
\eea
Remember also that the spin connection $\o_m{}^{cd}$ is a composite field of the vielbein, the gravitini, and the dilatation 
connection, eq.~\eqref{composite-Lorentz}.

We stress that the transformations 
\eqref{transf-standard-Weyl} 
form an algebra that
closes off-shell on a local extension of SU$(2,2|2)$. 
We will not need the explicit form of the algebra here, 
though it can be straightforwardly derived 
from results of \cite{deWit:1979dzm} and 
\cite{Butter_superspace:2011sr,Butter_reduction:2012xg}.
To conclude this subsection, 
for convenience, we include Table \ref{chiral-dilatation-weights} 
which summarises the non-trivial chiral and dilatation
weights of the fields and local gauge parameters 
of the standard Weyl multiplet. 
\begin{table}[hbt!]
\begin{center}
\begin{tabular}{ |c||c| c| c| c| c| c| c| c|c|c|c|} 
 \hline
& $e_m{}^a$
&$\psi_m{}_i$, $\xi_i$
&  $\bar\psi_m{}^i$, $\bar{\xi}^i$
& $\phi_m{}^i$, $\eta^i$ 
& $\bar\phi_m{}_i$, $\bar\eta_i$ 
&$\mathfrak{f}_{m}{}_c$
& $W^+_{ab}$
& $W^-_{ab}$ & 
$\Sigma^{i}$ 
& $\bar\Sigma_{i}$
&$D$
\\ 
\hline
 \hline
$\mathbb{D}$
&$-1$
&$-1/2$
&$-1/2$
&$1/2$
&$1/2$
&$1$
&$1$
&$1$
&$3/2$
&$3/2$
&$2$
\\
$Y$
&$0$
&$-1$
&$1$
&$1$
&$-1$
&$0$
&$-2$
&$2$
&$-1$
&$1$
&$0$
\\
\hline
\end{tabular}
\caption{\footnotesize{Summary of the non-trivial 
dilatation and chiral weights in the standard Weyl 
multiplet.}\label{chiral-dilatation-weights}}
\end{center}
\end{table}

\subsection{On-shell hypermultiplet 
and hyper-dilaton Weyl multiplet}
\label{hyper+HDWM}

A single on-shell hypermultiplet comprises $4+4$ degrees of freedom
described by a Lorentz scalar field $q^{i\underline{i}}$ 
and spinor fields 
$(\rho_\a^{\underline{i}}\,,\bar{\rho}^\ad_{\underline{i}})$
---
see \cite{Fayet:1975yi,FS2,deWit:1980gt,deWit:1980lyi,deWit:1983xhu,deWit:1984rvr}
together with \cite{deWit:1999fp,freedman,Lauria:2020rhc}
and references therein for superconformal approaches
to systems of on-shell hypermultiplets. 
The index $\underline{i}=\underline{1},\underline{2}$ is an SU(2) flavour 
index and the fields satisfy the following reality conditions
\bea
(q^{i\underline{i}})^*=q_{i\underline{i}}
~,~~~~~~
(\rho_\a^{\underline{i}})^*=\bar{\rho}_{\ad \underline{i}}
~,
\eea
together with the following dilatation and
chiral weight identities
\bsubeq
\bea
\mathbb{D}q^{i\underline{i}}
=q^{i\underline{i}}
~,~~~
\mathbb{D}\rho_\a^{\underline{i}}
&=&\frac{3}{2}\rho_\a^{\underline{i}}
~,~~~
\mathbb{D}\bar{\rho}_{\ad\underline{i}}
=\frac{3}{2}\bar{\rho}_{\ad\underline{i}}
~,
\\
Yq^{i\underline{i}}
=0
~,~~~
Y\rho_\a^{\underline{i}}
&=&\rho_\a^{\underline{i}}
~,~~~
Y\bar{\rho}_{\ad\underline{i}}
=-\bar{\rho}_{\ad\underline{i}}
~.
\eea
\esubeq
The multiplet, which has the field $q^{i\underline{i}}$ 
as its superconformal primary,
is characterised by the following local superconformal transformations
\cite{deWit:1980gt,deWit:1980lyi,deWit:1983xhu,deWit:1984rvr,deWit:1999fp,freedman,Lauria:2020rhc}
\bsubeq\label{hyper-susy}
\bea
\delta q^{i\underline{i}} 
&=& 
\frac{1}{2}\xi^{ i}\rho^{\underline{i}}
-\frac{1}{2}\bar{\xi}^{i} \bar{\rho}^{\underline{i}}
+\lambda^{i}{}_{k}q^{k\underline{i}} 
+ \lambda_{\mathbb{D}}q^{i\underline{i}}
~,
\label{d-qii}
\\
\delta \rho_{\a}^{\underline{i}} 
&=&
-4\ri(\sigma^a\bar{\xi}_k)_\a 
\nabla_a q^{k\underline{i}}
 +\hf \lambda_{ab} (\s^{ab} \rho^{\underline{i}})_{\a}    
 + \ri\lambda_Y\rho_{\a}^{\underline{i}}
+ \frac{3}{2}\lambda_{\mathbb{D}}\rho_{\a}^{\underline{i}} 
+ 8\eta_\a^k q_k{}^{\underline{i}}
~,
\label{d-rho}
\\
\delta \bar{\rho}^{\ad}_{\underline{i}} 
&=&
~\,\,\,4\ri(\tilde{\sigma}^a{\xi}^k)^{\ad}\nabla_a q_{k{\underline{i}}}
 +\hf \lambda_{ab} (\tilde{\s}^{ab}\bar{\rho}_{\underline{i}} )^{\ad}  
 - \ri\lambda_Y\bar{\rho}^{\ad}_{\underline{i}}
+ \frac{3}{2}\lambda_{\mathbb{D}}\bar{\rho}^{\ad}_{\underline{i}} 
- 8\bar{\eta}^\ad_k q^k{}_{\underline{i}}
~,
\label{d-b-rho}
\eea
\esubeq
where 
\bea
\nabla_aq^{i\underline{i}}
=
\cD_a q^{i\underline{i}}
 -\frac{1}{4}\psi_a{}^{i}
 \rho^{\underline{i}}
 + \frac{1}{4}\bar{\psi}_a{}^{i} 
 \bar{\rho}^{\underline{i}}
 ~.
 \label{DD'q}
\eea
In contrast with the standard Weyl 
multiplet described in the previous 
subsection,
the algebra of the local
transformations \eqref{hyper-susy} 
closes 
only when equations of motion for the 
fields are imposed,
see for example \cite{deWit:1999fp,Lauria:2020rhc} for a detailed analysis.
In our notations, the covariant equations of motion
of $q^{i\underline{i}}$
and $(\rho_\a^{\underline{i}},\,\bar{\rho}^{\ad}_{\underline{i}})$
are:
\bsubeq\label{on-shell-hyper}
\bea
\left(\nabla_{a}
\rho^{\underline{i}}\,
\s^a\right)_{\dot{\alpha}}
&=&
~\,\,\, \frac{\ri}{2}
(\bar{\rho}^{\underline{i}}\,
\tilde{\s}^{cd})_{\ad}
W^-_{cd}
+6\ri\bar{\Sigma}_{\dot{\alpha} k}q^{k\underline{i}}
 ~,\\
\left(
\nabla_a
\bar{\rho}_{\underline{i}}\,
\tilde{\s}^a\right)^{\alpha}
&=& 
-\frac{\ri}{2}
(\rho_{\underline{i}}\,\s^{cd})^{\alpha}
W^+_{cd}
+6\ri\Sigma^{\a k}q_{k\underline{i}}
~,\\
\Box q^{i\underline{i}} 
&=& 
 -\frac{3}{2}Dq^{i\underline{i}} 
 ~,~~~~~~
\Box:=\de^a\de_a
~.
\label{boxq-0}
\eea
\esubeq
The expressions for 
$\nabla_{a}\rho_{\alpha}^{\underline{i}}$, 
$\nabla_{a}\bar{\rho}^{\dot{\alpha}}_{\underline{i}}$, 
and $\square q^{i\underline{i}}$ in terms of the derivatives 
$\cD_a$ are given by
\bsubeq
\bea
\nabla_a \rho_\a^{\underline{i}}
&=& 
\cD_a\rho_\a^{\underline{i}} 
+2\ri{(\sigma^b{\bar{\psi}_a}{}_k)}_{\a}\left( 
\cD_b q^{k\underline{i}} 
-\frac{1}{4}{\psi_b}^{k}
\rho^{\underline{i}} 
+\frac{1}{4}{\bar{\psi}_b}{}^{k}
\bar{\rho}^{\underline{i}}
\right) 
+4{\phi_a}_{\a k}q^{k\underline{i}}
~,
\\\nabla_a \bar{\rho}^\ad_{\underline{i}}
&=& 
\cD_a\bar{\rho}^\ad_{\underline{i}} 
-2\ri{(\tilde{\sigma}^b{{\psi}_a}^{{ k}})}^{\dot{\a}}\left( 
\cD_b q_{k\underline{i}} 
- \frac{1}{4}
{\psi_b}_k\rho_{{\underline{i}}} 
+\frac{1}{4}
{\bar{\psi}_b}{}_k\bar{\rho}_{ {\underline{i}}}
\right) 
- 4{\bar{\phi}_a}{}^{\ad k}q_{k{\underline{i}}}
~,
\\
\square q^{i\underline{i}} 
&=& 
\cD^a\cD_a q^{i\underline{i}}  
- 2 {\mathfrak{f}_a}^a  q^{i \underline{i}}
- \frac{1}{4}\rho^{\underline{i}} \cD_a{\psi^a}^{i}
+ \frac{1}{4}\bar{\rho}^{\underline{i}} \cD_a {\bar{\psi}^a}{}^{ i} 
- \frac{1}{2}{\psi^a}^{ i}\cD_a\rho^{\underline{i}}
+ \frac{1}{2} {\bar{\psi}^a}{}^{ i}\cD_a\bar{\rho}^{\underline{i}}
\nonumber\\
&&
- \frac{\ri}{4}\phi_a{}^i \sigma^a \bar{\rho}^{\underline{i}} 
+ \frac{\ri}{4}\bar{\phi}_a^i \tilde{\sigma}^a \rho^{\underline{i}} 
+\ri
({\psi_a}^{(i}\sigma^b {\bar{\psi}^{k)a}})
\cD_b q_k{}^{\underline{i}} 
+ \frac{3\ri}{4}({\psi_a}^{ i} \sigma^a \bar{\Sigma}_{l}) q^{l \underline{i}} 
+\frac{3\ri}{4} ({\bar{\psi}}_a^i \tilde{\sigma}^a\Sigma_l)q^{l\underline{i}} 
\non\\
&&
-\frac{\ri}{16}  ({\bar{\psi}}_a{}^i 
\tilde{\sigma}^a\s^{cd} 
\rho^{\underline{i}})
W^+_{cd}
- \frac{\ri}{16} ({\psi_a}^{ i}\sigma^a \tilde{\s}^{cd} \bar{\rho}^{\underline{i}}) W^-_{cd}
-({\psi_a}^{i}{\phi^a}{}_{k})
q^{k\underline{i}}
+ ({\bar{\psi}}_a{}^i \bar{\phi}^a{}_k)q^{k\underline{i}} 
 \nonumber \\
&& 
- \frac{\ri}{4}
({\psi_a}^{(i}
\sigma^b
{\bar{\psi}^{k)}{}^a })
({\psi_{b}{}_{k}}\rho^{\underline{i}}) 
+ \frac{\ri}{4}({\psi_a}^{(i}\sigma^c{\bar{\psi}^{k)}}{}^{a})({\bar{\psi}_c}{}_{k}\bar{\rho}^{\underline{i}}) ~.
\eea
\esubeq

It is important to stress that equations \eqref{on-shell-hyper}
are typically read as equations of motion for the hypermultiplet fields, see e.\,g., \cite{deWit:1980gt,deWit:1980lyi,deWit:1983xhu,deWit:1984rvr,deWit:1999fp,freedman,Lauria:2020rhc}. 
They certainly are dynamical equations for 
 $q^{i\underline{i}}$ and $(\rho_\a^{\underline{i}}\,,\bar{\rho}^\ad_{\underline{i}})$
in a flat background (with no central charges as in our case) 
where all conformal supergravity fields are set to zero
\cite{Fayet:1975yi,FS2}. 
For this reason, 
the multiplet is typically referred to as the on-shell 
hypermultiplet. However, such an interpretation is not necessary in a curved background described by the 
standard Weyl multiplet. 
In fact, 
the equations \eqref{on-shell-hyper} can be interpreted as algebraic equations for the 
standard Weyl multiplet that determine
the fields $(\S^{\a i},\bar{\S}_{\ad i})$ and $D$ in terms of 
$q^{i\underline{i}}$ and $(\rho_\a^{\underline{i}}\,,\bar{\rho}^\ad_{\underline{i}})$
together with the other independent fields of the standard Weyl multiplet.
If we assume that $q^{i\underline{i}}$ is an invertible matrix, which is equivalent to imposing
\bea
q^2:=q^{i\underline{i}}q_{i\underline{i}}=\ve_{ij}\ve_{\underline{i}\underline{j}}q^{i\underline{i}}q^{j\underline{j}}
=2\det{q^{i\underline{i}}}\ne0
~,
\eea
then the following relations hold
\bsubeq\label{compositeSSbD}
\bea
{\Sigma}^{{\alpha} i} 
&=&
2q^{-2}q^{i\underline{i}}\Bigg[
- \frac{\ri}{2}
(\cD_a {\bar{\rho}}_{\underline{i}}
\,\tsigma^a)^\a
 +(\psi_a{}^j \s^b \tilde{\s}^a)^\a \left( 
 \cD_b q_{j\underline{i}} 
 -\frac{1}{4}{{\psi}_b}{}_j
 \rho_{ \underline{i}}
 +\frac{1}{4}
 {\bar\psi_b}{}_{j}\bar\rho_{\underline{i}}
 \right)
 \non\\
&&~~~~~~~~~\,\,
+ \frac{2}{3}
\big(\Psi_{ab}{}_j \s^{ab}\big)^{\a}
q^j{}_{\underline{i}} 
+\frac{1}{4}
(\rho_{\underline{i}}\s^{cd})^{\a}
W^+_{cd} 
+\frac{\ri}{6}
({\bar{\psi}}_a{}_{j}
\ts^a\s^{cd})^{\a} 
q^j{}_{\underline{i}} W^+_{cd}
\,\Bigg] 
~,
\label{sigma}
\\
\bar{\Sigma}_{\dot{\alpha} i} 
&=&
2q^{-2}q_{i\underline{i}}\Bigg[
- \frac{\ri}{2}
(\cD_a \rho^{\underline{i}}
\,\sigma^a)_\ad 
 -({\bar{\psi}_a}{}_j \tilde{\s}^b \sigma^a)_\ad \left( 
 \cD_b q^{j\underline{i}} 
 - \frac{1}{4}
 {\psi_b}{}^{j}\rho^{\underline{i}}
 +\frac{1}{4}
 {\bar{\psi}_b}{}^{j}
 \bar{\rho}^{\underline{i}}\right)\nonumber \\
&&~~~~~~~~~~
-\frac{2}{3}
\big(
\overline{\Psi}_{ab}{}^j \ts^{ab}
\big)_{\ad}
q_j{}^{\underline{i}} 
-\frac{1}{4}
(\bar\rho^{\underline{i}}\ts^{cd})_{\ad}
W^-_{cd} 
+\frac{\ri}{6}
(\psi_a{}^{j}
\s^a\ts^{cd})_{\ad} 
q_j{}^{\underline{i}} W^-_{cd}
\,
\Bigg] 
~,
\label{sigma-bar}
\\
D 
&=& 
q^{-2}q_{i\underline{i}}\Bigg[
~\cD^a\cD_a q^{i\underline{i}}  
+\frac{1}{6}R\,q^{i\underline{i}} 
-\frac{\ri}{8}({\bar{\psi}}_a{}^i \tilde{\sigma}^a\s^{cd}
\rho^{ \underline{i}})W^+_{cd}
- \frac{\ri}{2}\phi_a{}^i \sigma^a \bar{\rho}^{\underline{i}} 
- \frac{1}{2}\rho^{\underline{i}} \cD_a{\psi^a}^{i}
\nonumber\\
&&
~~~~~~~~\,\,
- {\psi^a}^{ i}\cD_a\rho^{\underline{i}}
+ 2 ({\psi_a}^{i}{\phi^a}^{j})
{q_j}^{\underline{i}}
+ \frac{3\ri}{2}
({\psi_a}^{ i} \sigma^a \bar{\Sigma}_{j}) 
q^{j \underline{i}}
+ \frac{\ri}{2}({\psi_{aj}}\sigma^a \bar{\Sigma}^j)q^{i\underline{i}} 
\nonumber \\
&&~~~~~~~~~
+ \frac{1}{6}\varepsilon^{mnpq}({{\bar{\psi}_{m}}^j} \tilde{\sigma}_n\cD_p\psi_{qj})q^{i\underline{i}}
+ \frac{1}{3}W^{ab+}({{\bar{\psi}_a}^j}{\bar{\psi}_{bj}})q^{i\underline{i}}
+\ri({\psi_a}^{(i}\sigma^b {\bar{\psi}^{j)a}}) \cD_b q_j{}^{\underline{i}} \nonumber \\
&&~~~~~~~~~
- \frac{\ri}{2} ({\psi_a}^{(i}\sigma^b{\bar{\psi}^{j)}{}^a }) ({\psi_{b}{}_{j}}\rho^{\underline{i}})\Bigg] 
+ \text{c.c.}
~.
\label{D_def}
\eea
\esubeq
In the expression for $D$, 
eq.~\eqref{D_def},
remember that
$(\S^i,\bar\S_i)$
and 
$(\phi^i,\bar\phi_i)$, together with the 
spin connection $\o_m{}^{cd}$,
are composite fields.
Note that so far we have only used one of the four 
equations that are equivalent
to \eqref{boxq-0} to solve for $D$ in eq.~\eqref{D_def}.
It is simple to show that the remaining independent
three equations are equivalent to the following 
\be
\de^a (q^{i(\underline{i}}\de_aq_i{}^{\underline{j})})=0
~.
\label{2-4-constraints}
\ee
As we are going to explain in detail below, 
this equation is solved by turning 
the SU(2)$_R$ connection $\phi_m{}^{kl}$
into a composite field.

As a next step in the construction of the hyper-dilaton Weyl multiplet, we note that, accompanied to an on-shell hypermultiplet
there is always a triplet of composite linear multiplets
\cite{FS2,deWit:1980gt,deWit:1980lyi,deWit:1983xhu}. 
An $\cN=2$ off-shell linear multiplet  
\cite{N=2tensor,Siegel:1978yi,Siegel80,SSW,deWit:1980lyi,deWit:1982na,KLR,LR3} 
comprises the following covariant fields: 
an SU$(2)_R$ triplet of Lorentz scalar fields $G^{ij}$ subject to the reality condition $(G^{ij})^*=G_{ij}$; 
spinor fields $(\chi_{\a i},\,\bar{\chi}^{\ad i})$;
a complex scalar field $(F,\,\bar{F})$;
and a covariant real closed anti-symmetric three-form $H_{abc}$, which is equivalent to a conserved dual vector 
$\tilde{H}^a:=\frac{1}{6}\ve^{abcd}H_{bcd}$. 
Their local superconformal transformations in a standard Weyl multiplet background
are given by \cite{deWit:1980lyi}
\bsubeq
\label{linear-multiplet}
\bea
\d G_{ij}
&=&
2 \xi_{(i} \chi_{j)}
+ 2\bar\xi_{ (i}  \bar{\chi}_{j )}
- 2\lambda_{(i}{}^{k}  G_{j) k} 
+2 \lambda_{\mathbb{D}} G_{ i j}
~,
\\
\d \chi_{\a i}
&=&
-\xi_{\a i} F 
-4\ri\tilde{H}_a (\s^a \bar{\xi}_{i})_\a 
+ \ri   
(\s^a \bar{\xi}^{j})_\a 
\de_{a} G_{i j}
+ 4 \eta_{\a}^{j}   G_{j i}
\nonumber\\ 
&& 
+\frac{1}{2} \lambda^{a b}
(\s_{a b}\chi_{i})_{\a} 
-\lambda_i{}^j \chi_{j \a}
+\frac{5}{2}\lambda_{\mathbb{D}}\chi_{\a i} +\ri \lambda_{Y} \chi_{\a i} 
~,
\\
\d \bar{\chi}^{\ad i}
&=&
-\bar\xi^{\ad i} \bar{F}
+4\ri\tilde{H}_a (\ts^a {\xi}^{i})^\ad 
+\ri(\ts^a \xi_{j})^\ad 
\de_{a} G^{i j}
+ 4 \bar\eta^{\ad}_{j}G^{j i}
\nonumber\\ 
&& 
+\frac{1}{2} \lambda^{a b}
(\ts_{a b}\bar\chi^{i})^{\ad} 
+\lambda^i{}_j \bar{\chi}^{j\ad}
+\frac{5}{2}\lambda_{\mathbb{D}}
\bar\chi^{\ad i} 
-\ri \lambda_{Y} \bar\chi^{\ad i} 
~,
\\
\d F
&=&
- 2 \ri \bar{\xi}^{i} \tilde{\s}^a \de_a \chi_i
+(\bar{\xi}^{i} \ts^{cd}\bar{\chi}_i)
W^-_{cd}
- 6 (\bar{\xi}^{i}\bar{\Sigma}^{ j})
G_{i j} 
+ 4\eta^{i} \chi_{i}
\nonumber\\
&&
+3 \lambda_{\mathbb{D}} F 
+ 2 \ri \lambda_{Y} F 
~,\\
\d \bar{F}
&=&
- 2 \ri{\xi}_{i}\s^a \de_a \bar\chi^i
-({\xi}_{i} \s^{cd}\chi^{i})W^+_{cd}
+6(\xi^{i}\Sigma^{j})G_{i j} 
+4\bar\eta_{i} \bar\chi^{i}
\nonumber\\
&&
+3\lambda_{\mathbb{D}} \bar{F} 
- 2 \ri \lambda_{Y} \bar{F} 
~,
\\
\d \tilde{H}_{a}
&=&
\frac{1}{2} \xi_{i} \s_{a b} \de^b \chi^{ i}
-\frac{\ri}{16} 
(\xi_{i} \s_a\ts^{cd} \bar{\chi}^{i})
W^-_{cd}
- \frac{3\ri}{8} 
(\xi_{i} \s_a\bar{\Sigma}_j) 
G^{ij} 
\non\\
&&
-\frac{1}{2}\bar{\xi}^{i}\tilde{\s}_{a b} \de^b \bar{\chi}_i
- \frac{\ri}{16} 
(\bar\xi^{i}\tilde{\s}_a\s^{cd}\chi_{i})
W^+_{cd}
-\frac{3\ri}{8}
(\bar\xi^{i}\tilde{\s}_a\Sigma^{j}) 
G_{ij}
\non\\
&&
+ \lambda_a{}^b\tilde{H}_b
+3\lambda_{\mathbb{D}} \tilde{H}_a 
-\frac{3\ri}{4} 
\eta^{i}\s_a \bar{\chi}_i
+\frac{3 \ri}{4}
\bar{\eta}_{i} \tilde{\s}_a \chi^i
~,
\eea
\esubeq
where
\bsubeq
\bea
\nabla_{a} G_{i j}
&=&
\cD_{a} G_{i j} - \psi_{a (i} \chi_{j)} -\bar{\psi}_{a (i} \bar{ \chi}_{j)} 
~,
\\
\nabla_{a} \chi_{\a i}
&=&
\cD_{a} \chi_{\a i} 
+\frac{1}{2} \psi_{a \a i}  F 
+2\ri (\s^b \bar{\psi}_{a i})_\a   
\tilde{H}_b 
-\frac{\ri}{2}
(\s^b \bar{\psi}_{a}{}^{j})_\a
\de_{b}G_{ij} 
-2 \phi_a{}_\a^j  G_{i j}
~,
\\
\nabla_{a} \bar\chi^{\ad i}
&=&
\cD_{a}\bar\chi^{\ad i} 
+\frac{1}{2}\bar\psi_{a}{}^{ \ad i}  
\bar F 
-2\ri (\ts^b {\psi}_{a}{}^{i})_\a   
\tilde{H}_b 
-\frac{\ri}{2}
(\ts^b\psi_{a}{}_{j})^\ad
\de_{b}G^{ij} 
-2 \bar\phi_a{}^\ad_j  G^{i j}
~.
\eea
\esubeq
The covariant conservation equation for $\tilde{H}_a$ is
\bea
\nabla^a \tilde{H}_{a} 
&=& 
\frac{3}{8}
\S^{i}\chi_{i}
+\frac{3}{8}
\bar{\S}_{i}\bar{\chi}^{i}
~.
\label{covariant-current}
\eea
The constraint implies the existence of a gauge two-form potential, $b_{mn}=-b_{nm}$, 
and its exterior derivative $h_{mnp}:=3\pa_{[m}b_{np]}$. The solution of \eqref{covariant-current} is
\bea
\tilde{H}_a 
=  \frac{1}{6} \ve_a{}^{bcd} \Big(h_{bcd}
-\frac{3 \ri}{4}\psi_b{}_i \s_{cd}\chi^i 
-\frac{3 \ri}{4}
\psib_b{}^i\tilde{\s}_{cd}\bar{\chi}_i
-\frac{3}{4} (\psi_b{}^i\s_c\psib_d{}^j) G_{ij} \Big)
~,
\label{solution-H}
\eea
where
$h_{abc}=e_a{}^me_b{}^ne_c{}^ph_{mnp}$.
The locally superconformal transformations of $b_{mn}$ are 
\bea
\d b_{mn}
&=&
\frac{\ri}{2} \xi_i\s_{mn} \chi^i
+ \frac{\ri}{2}
\bar{\xi}^i \tilde{\s}_{m n} \bar{\chi}_i  +\frac{1}{2}\Big(
\psib_{[m}{}^i\s_{n]}\xi^j
-\psi_{[m}{}^i\s_{n]}\bar{\xi}^j\Big)
G_{ij}
+2\pa_{[m}l_{n]}
~,
~~~~~~
\label{d-bmn}
\eea
where we have also included 
the vector gauge transformation
$\d_l b_{mn}=2\pa_{[m}l_{n]}$
that leaves 
$h_{mnp}$
and
$\tilde{H}^a$ invariant.
For convenience, we have summarised the dilatation and chiral weights
of the fields of the linear multiplet in Table 
\ref{chiral-dilatation-weights-LINEAR}.
\begin{table}[hbt!]
\begin{center}
\begin{tabular}{ |c||c| c| c| c| c| c| c|} 
 \hline
& $G_{ij}$ 
& $\chi_{\a i}$
&$\bar\chi^{\ad i}$
& $F$
& $\bar{F}$
& $\tilde{H}^a$
&$b_{mn}$
\\ 
\hline
 \hline
$\mathbb{D}$
&$2$
&$5/2$
&$5/2$
&$3$
&$3$
&$3$
&$0$
\\
$Y$
&$0$
&$1$
&$-1$
&$2$
&$-2$
&$0$
&$0$
\\
\hline
\end{tabular}
\caption{\footnotesize{Summary of the 
dilatation and chiral weights in the off-shell linear multiplet.}
\label{chiral-dilatation-weights-LINEAR}}
\end{center}
\end{table}

Now that we have reviewed the structure 
of a locally superconformal linear multiplet, 
a straightforward analysis shows that,
assuming $q^{i\underline{i}}$ and $(\rho_\a^{\underline{i}}\,,\bar{\rho}^\ad_{\underline{i}})$ describe an on-shell 
hypermultiplet in a standard Weyl multiplet background with transformation rules \eqref{hyper-susy},
the following composite fields define a triplet of linear multiplets
\cite{deWit:1984rvr}
\bsubeq\label{composite-linear}
\bea
G_{ij}{}^{\underline{i}\underline{j}}
&=&
q_{(i}{}^{\underline{i}}q_{j)}{}^{\underline{j}}
=q_{i}{}^{(\underline{i}}q_{j}{}^{\underline{j})}
~,
~~~~~~~~~~~~~~~~~~~~~~~~~
(G_{ij}{}^{\underline{i}\underline{j}})^*
=
G^{ij}{}_{\underline{i}\underline{j}}
~,\\
\chi_{\alpha i}{}^{\underline{i}\underline{j}}
&=& 
\frac{1}{2} q_{i}{}^{(\underline{i}}\rho_{\alpha}^{\underline{j})}
~,
~~~
\bar\chi^{\ad i}{}_{\underline{i}\underline{j}}
=
-\frac{1}{2} q^{i}{}_{(\underline{i}}\bar{\rho}^{\ad}_{\underline{j})}
~,
~~~~~~~~~
(\chi_{\alpha i}{}^{\underline{i}\underline{j}})^*
=
\bar\chi_\ad^i{}_{\underline{i}\underline{j}}
~,\\
F^{\underline{i}\underline{j}} 
&=& 
\frac{1}{8}\rho^{(\underline{i}}\rho^{\underline{j})}
~,~~~
\bar{F}_{\underline{i}\underline{j}} 
= 
\frac{1}{8}\bar\rho_{(\underline{i}}
\bar\rho_{\underline{j})}
~,~~~~~~~~~~~~~~~~~
(F^{\underline{i}\underline{j}})^*
=
\bar F_{\underline{i}\underline{j}}
~,
\label{composite-F}
\\
\tilde{H}^a{}^{\underline{i}\underline{j}}
&=& -\frac{1}{4}q^{i(\underline{i}} \nabla^a{q_i}^{\underline{j})}
+\frac{\ri}{32}
\rho^{(\underline{i}}\sigma^a\bar{\rho}^{\underline{j})}
~,~~~~~~~~~~~~~
(\tilde{H}^a{}^{\underline{i}\underline{j}})^*
=\tilde{H}^a{}_{\underline{i}\underline{j}}
~.
\label{linear-H}
\eea
\esubeq
These fields all transform according to 
\eqref{linear-multiplet} and
 each of the previous fields is symmetric in $\underline{i}$ and $\underline{j}$.
Within the previous composite fields, the field 
$\tilde{H}^a{}^{\underline{i}\underline{j}}$ 
is particularly interesting.
In fact, equation \eqref{linear-H}
together with \eqref{solution-H}
represent the solution to the
constraint \eqref{2-4-constraints} 
and can be used to express the 
${\rm SU}(2)_R$ connection 
$\phi_m{}^{ij}$ as a composite field.
By introducing the derivative 
\bea
\mathbf{D}_a 
&=& 
{e_a}^m\left( \partial_m - \frac{1}{2}{\omega_m}^{cd}M_{cd} 
- \ri A_m Y 
- b_m\mathbb{D}\right) 
= \cD_a + {e_a}^m{\phi_m}^{ij}J_{ij}
~,
\eea
and by using \eqref{DD'q},
eq.~\eqref{linear-H} can be rearranged for the SU(2)$_R$ gauge 
connection as follows
\bea
 \phi_a{}^{ij}  
 &=& 
 4q^{-4}q^{(i}{}_{\underline{i}}q^{j)}{}_{\underline{j}}
 \Bigg[\, 
 q^{k\underline{i}} \mathbf{D}_a q_k{}^{\underline{j}}
 - \frac{1}{4}q^{k\underline{i}} 
 ({\psi_a}{}_k \rho^{\underline{j}}) 
 + \frac{1}{4}q^{k\underline{i}}
 ({\bar{\psi}_a}{}_k \bar{\rho}^{\underline{j}})
 - \frac{\ri}{8}
 \rho^{\underline{i}}
 \s_a
 \bar{\rho}^{\underline{j}}
 +4{\tilde{H}_a{}^{\underline{i}\underline{j}}}\,\Bigg] 
 ~.~~~~~~~~~
 \label{compositeSU2}
\eea

This concludes the definition of the hyper-dilaton Weyl multiplet. The final result of our analysis is that we have identified
a new representation of the off-shell local 4D, $\cN=2$ superconformal algebra in terms of the following independent 
fields: 
${e_m}^a$, $b_m$, $A_{m}$, $W_{ab}$, $q^{i\underline{i}}$, $b_{mn}{}^{\underline{i}\underline{j}}$,
$(\psi_m{}_i,\bar\psi_m{}^i)$, and $(\rho^{\underline{i}},\bar\rho_{\underline{i}})$. The multiplet has precisely the same number of off-shell
degrees of freedom as the standard Weyl multiplet,
$24+24$. 
Table \ref{dof2} summarises the counting of degrees of freedom, 
underlining the symmetries acting on the fields. 
\begin{table}[hbt!]
\begin{center}
\begin{tabular}{ |c c c c c c c c |c c c c|} 
 \hline
${e_m}^a$ & ${\omega_m}^{ab}$ & $b_m$ & ${\mathfrak{f}_m}{}_a$ & $\phi_{m}{}^{ij}$ & $A_{m}$ & $\psi_m{}_i$ & $\phi_m{}^i$ & $W_{ab}$ & $\rho^{\underline{i}}$ & $q^{i\underline{i}}$ & $b_{mn}{}^{\underline{i}\underline{j}}$\\ 
$16B$ & $0$ & $4B$ & $0$ & $0$ & $4B$ & $32F$ & $0$ & $6B$ & $8F$ & $4B$ & $18B$\\
\hline
$P_a$ & $M_{ab}$ & $\mathbb D$ & $K_a$ & $J^{ij}$ & $Y$ & $Q$ & $S$ & {} & ${}$ & ${}$ & $\lambda_m{}^{\underline{i}\underline{j}}$-sym\\
$-4B$ & $-6B$ & $-1B$ & $-4B$ & $-3B$ & $-1B$ & $-8F$ & $-8F$ & {} & {} & {} & $-9B$\\
\hline
\multicolumn{12}{|c|}{Result: $24+24$ degrees of freedom}\\
\hline
\end{tabular}
\caption{\footnotesize{Degrees of freedom and symmetries of the hyper-dilaton Weyl multiplet. Row one gives all the fields in the multiplet. Row two gives the number of independent components of these fields -- composite connections are counted with zero degrees of freedom. 
Row three gives the gauge symmetries. Note that the parameter $\lambda_m{}^{\underline{i}\underline{j}}$ 
describes the vector symmetry associated with the gauge two-forms
$b_{mn}{}^{\underline{i}\underline{j}}$ 
with field strength
three-forms $h_{mnp}{}^{\underline{i}\underline{j}}$ 
and $\tilde{H}^a{}^{\underline{i}\underline{j}}$.
Row four gives the number of gauge degrees of freedom to be subtracted when counting the total degrees of freedom. 
Row five gives the resulting number of degrees of freedom. }\label{dof2}}
\end{center}
\end{table}
Note that with the ingredients provided so far, it is 
a straightforward
exercise to obtain
the locally superconformal 
transformations of the fundamental fields
of the hyper-dilaton Weyl multiplet
written only in terms of fundamental fields.
These are given by
\eqref{d-vielbein}--\eqref{d-b-gravitino},
\eqref{d-U1}--\eqref{d-W},
\eqref{d-bmn}, and
\eqref{d-qii}--\eqref{d-b-rho}
after using the appropriate identities 
for all the composite fields 
$\o_m{}^{cd}$, 
${\mathfrak{f}_m}{}_a$,
$\phi_{m}{}^{ij}$, 
$(\phi_{m}{}_{i},\bar\phi_{m}{}^{i})$, 
$(\Sigma^{\a i},\bar\Sigma_{\ad i})$,
and
$D$
respectively given by eqs.~\eqref{composite-Lorentz},
\eqref{compositeSU2},
\eqref{composite-S-connection},
and
\eqref{compositeSSbD}.

It is important to underline 
that the local gauge transformations of the 
hyper-dilaton Weyl 
multiplet form an algebra that
closes off-shell on a local extension of SU$(2,2|2)$. 
In fact, by construction the resulting algebra
is identical to the one of the standard Weyl multiplet transformations
\eqref{transf-standard-Weyl} 
(see \cite{deWit:1979dzm} and 
\cite{Butter_superspace:2011sr,Butter_reduction:2012xg}
for detail on the local algebra),
with the only important subtlety being that
the structure functions will have more composite fields.


\section{Gauge fixing and M\"uller's 
Poincar\'e supergravity}
\label{section-3}

As explained in the introduction, one of the motivations of our analysis was to show that 
the $32+32$ off-shell multiplet of 4D, $\cN=2$ 
Poincar\'e supergravity constructed by M\"uller in \cite{Muller_hyper:1986ts}
could be derived by superconformal techniques starting from the hyper-dilaton Weyl multiplet. In this section
we explain how this goes. We first focus on the structure of the multiplet and then explain
how to construct the Poincar\'e supergravity
action derived in \cite{Muller_hyper:1986ts}.
At the end of this section
we also extend the results of \cite{Muller_hyper:1986ts}
by adding a new $BF$-coupling which induces 
a scalar potential for the dilaton without a standard 
$R$-symmetry gauging.

\subsection{Hyper-Dilaton Poincar\'e supergravity multiplet}

To recover a multiplet of Poincar\'{e} supergravity, compensating multiplets must be coupled to the 
off-shell conformal supergravity multiplet to fix some of the local 
superconformal symmetries
--- see \cite{freedman,Lauria:2020rhc} for reviews. 
Below we will describe how to recover the multiplet described in \cite{Muller_hyper:1986ts}
which we denote as the hyper-dilaton Poincar\'e multiplet.
The construction is straightforward. 
We simply need to couple the hyper-dilaton Weyl multiplet to a
single 
off-shell vector multiplet compensator and then appropriately 
gauge fix to eliminate all symmetries except
 local supersymmetry, Lorentz, and the vector gauge symmetry of the gauge two-forms
$b_{mn}{}^{\underline{i}\underline{j}}$.

It is straightforward to define an off-shell 4D, $\cN=2$ Abelian vector multiplet in a hyper-dilaton Weyl multiplet background. 
As a first step consider an Abelian vector multiplet 
\cite{Fayet:1975yi,Grimm:1977xp}
in a standard Weyl multiplet background
\cite{deWit:1979dzm,deWit:1984rvr,deWit:1984wbb,Cremmer:1984hj}. 
This is described by a complex scalar field $\phi$
and its conjugate $\bar{\phi}=(\phi)^*$, 
gaugini $(\lambda_\alpha^i,\bar\lambda^\ad_i)$
such that $(\lambda_\alpha^i)^*=\bar\lambda_{\ad i}$, 
a triplet of auxiliary fields $X^{ij}=X^{ji}$ 
satisfying the reality condition $(X^{ij})^*=X_{ij}$,
and a real Abelian gauge connection $v_m$ or, equivalently,
its covariant real field strength ${F}_{ab}$ given by 
\bea
 F_{ab} &=&
e_a{}^m e_b{}^n {f}_{mn}
- \frac{\ri}{2} \psi_{[a}{}_k \s_{b]}\bar{\l}^k
+ \frac{\ri}{2} \psib_{[a}{}^k\tilde{\s}_{b]}\l_k 
- \hf (\psi_{a}{}_k \psi_{b}{}^k) \bar{\phi}
+ \hf (\psib_{a}{}^k \psib_{b}{}_k) \phi 
~,~~~~~~
\label{covF}
\eea
where ${f}_{mn}=2\pa_{[m}v_{n]}$.
By construction $F_{ab} $ satisfies the Bianchi identity 
\bea
\de_{[a}F_{bc]} 
=
-\frac{ \ri}{2} R(Q)_{[ab}{}_j \s_{c]}\bar{\l}^{j} 
+\frac{ \ri}{2}R(\bar{Q})_{[ab}{}^j \ts_{c]}\l_{j}
~,
\eea
that is solved by \eqref{covF}. The non-trivial 
dilatation and chiral weights of the vector multiplet fields are summarised in Table
\ref{weights-vector}.
\begin{table}[hbt!]
\begin{center}
\begin{tabular}{ |c||c| c| c| c| c| c| c|} 
 \hline
& $\phi$ 
& $\bar{\phi}$
&$\lambda_\alpha^i$
&  $\bar{\lambda}^\ad_i$
& $X^{ij}$
& $F_{ab}$
&$v_m$
\\ 
\hline
 \hline
$\mathbb{D}$
&$1$
&$1$
&$3/2$
&$3/2$
&$2$
&$2$
&$0$
\\
$Y$
&$-2$
&$2$
&$-1$
&$1$
&$0$
&$0$
&$0$
\\
\hline
\end{tabular}
\caption{\footnotesize{Summary of the 
dilatation and chiral weights in the off-shell
Abelian vector multiplet.}\label{weights-vector}}
\end{center}
\end{table}

The transformation rules of the vector multiplet fields in a standard Weyl multiplet background are
\bsubeq
\bea
\d \phi
&=&
\xi_i \l^i + \l_\mathbb{D} \phi - 2 \ri \l_Y \phi
~,
\\ 
\d \bar{\phi}
&=&
\bar{\xi}^i \bar{\l}_i + \l_{\mathbb{D}} \bar{\phi} + 2 \ri \l_Y \bar{\phi}
~,
\\
\d\lambda_\alpha^i
&=&
2 (\s^{a b} \xi^i)_\a F_{a b} + (\s^{a b}\xi^i)_\a W^{+}_{a b} \bar{\phi} 
- \frac{1}{2} \xi_\a{}_j X^{ij} + 2 \ri (\s^a \bar{\xi}^i)_\a \de_a \phi \nonumber \\
&&+ \frac{1}{2}\l^{a b}(\s_{a b} \l^i)_\a  + \l^i{}_j \l^j_\a + \frac{3}{2} \l_{\mathbb{D}} \l^i_\a - \ri \l_Y \l^i_\a + 4 \eta^i_\a \phi
~,
\label{transf-lambda}
\\
\d\bar{\lambda}^\ad_i
&=&
-2 (\tilde{\s}^{a b} \bar{\xi}_i)^\ad F_{a b} - (\tilde{\s}^{a b}\bar{\xi}_i)^\ad W^{-}_{a b} \phi 
- \frac{1}{2} \bar{\xi}^\ad{}^j X_{ij} + 2 \ri (\tilde{\s}^a \xi_i)^\ad \de_a \bar{\phi} \nonumber \\
&&+ \frac{1}{2}\l^{a b}(\tilde{\s}_{a b} \bar{\l}_i)^\ad  - \l_i{}^j \bar{\l}_j^\ad + \frac{3}{2} \l_{\mathbb{D}} \bar{\l}_i^\ad+ \ri \l_Y \bar{\l}^\ad_i + 4 \bar{\eta}^\ad_i \bar{\phi}
~, 
\label{transf-lambda-bar}
\\
\d X^{ij}
&=& -4 \ri \xi^{(i} \s^a \de_a \bar{\lambda}^{j)} - 4 \ri \bar{\xi}^{(i} \tilde{\s}^a \de_a \lambda^{j)} 
+ 2 \lambda^{(i}{}_k X^{j)k} 
+ 2 \lambda_{\mathbb{D}} X^{ij}
~,
\\
\d F_{ab}
&=&
\Bigg[
-\ri \xi_k \s_{[a}\de_{b]}\bar{\lambda}^k 
+ 2\big(\xi_k R(Q)_{ab}{}^k\big) \bar{\phi} 
-\hf (\xi_k\lambda^k) W_{ab}^- 
+ 2 \eta^k \s_{ab} \lambda_k 
+ {\rm c.c.} \Bigg] 
\non\\
&& 
~~
+2 \lambda_{\mathbb{D}} F_{ab} 
- 2 \lambda_{[a}{}^{c} F_{b]c}
~,
\\
\d v_m
&=& (\xi_k \psi_{m}{}^k) \bar{\phi} 
-(\bar{\xi}^k \bar{\psi}_{m k}) \phi 
+\partial_m \l_V
~,
\label{dvm}
\eea
\esubeq
where
\bsubeq
\bea
\de_a \phi
&=&
\cD_a \phi - \frac{1}{2}\psi_a{}_i \l^i
~,
\\
\de_a \bar{\phi}
&=&
\cD_a \bar{\phi} 
- \frac{1}{2}\bar{\psi}_a{}^i \bar{\l}_i
~,
\\
\de_a \l^i_\a
&=&
\cD_{a} \l^i_\a 
-(\s^{cd}\psi_a{}^{i})_\a\Big( F^+_{cd}
+ \frac{1}{2}  W^+_{cd} \bar{\phi} \Big)
+\frac{1}{4} \psi_{a}{}_\a{}_j X^{ij} 
\non\\
&&
-\ri (\s^b \bar{\psi}_{a}{}^i)_\a \de_b \phi - 2 \phi_a{}_\a^i \phi
~,
\\
\de_a \bar\l_i^\ad
&=& 
\cD_{a} \bar{\l}^\ad_i 
+ (\ts^{cd}\bar{\psi}_a{}_i)^{\ad}\Big( 
F^-_{cd}
+\frac{1}{2} W^-_{cd} \phi\Big)
+\frac{1}{4} \bar{\psi}_a{}^{\ad j} X_{ij} 
\non\\ 
&& 
- \ri (\tilde{\s}^b \psi_a{}_i)^\ad \de_b \bar{\phi} - 2 \bar{\phi}_a{}^\ad_i \bar{\phi}
~,
\eea
\esubeq
and we have also included in \eqref{dvm}
the gauge field transformation
parametrised by the local real parameter $\l_V$.
The transformations of the vector multiplet
in a hyper-dilaton Weyl multiplet background are precisely the same with the only subtlety 
that one has to interpret several 
standard Weyl multiplet fields as composite of 
$q^{i\underline{i}}$, $(\rho_{\a}^{{\underline{i}}},\,\bar{\rho}^{\ad}_{{\underline{i}}})$, 
and $b_{mn}{}^{\underline{i}\underline{j}}$.

The compensating vector multiplet contains $8+8$ off-shell 
degrees of freedom. Once added to the 
hyper-dilaton Weyl multiplet
we obtain the right number of off-shell degrees of freedom,
$32+32$,
of the hyper-dilaton Poincar\'e multiplet \cite{Muller_hyper:1986ts}
but in a manifestly superconformal setting. 
We can then obtain the structure of 
the Poincar\'e multiplet,
including its local transformation rules,
after gauge fixing.

The first set of gauge fixing conditions are
\bsubeq\label{gauge-conditions}
\bea
&\phi=1
~,~~~
\bar{\phi}=1
~,
\label{fixing-phi}
\\
&
b_m=0
~.
\label{fixing-K}
\eea 
The condition \eqref{fixing-phi} fixes dilatation and U(1)$_R$ symmetries, while \eqref{fixing-K} 
fixes special conformal $K^a$ symmetry.
Next we impose
\bea
\lambda_\alpha^i=0~,~~~
\bar{\lambda}^\ad_i=0
~,
\label{S-susy-gauge}
\eea
which gauge fixes $S$-supersymmetry. A characterising feature of the hyper-dilaton Weyl multiplet is that 
it contains an SU(2)$_R$ compensator, 
the $q^{i\underline{i}}$ fields. 
As a last gauge fixing condition we then impose
\be
q^{i\underline{i}} =- \ve^{i\underline{i}} \re^{-U}
~~~\Longleftrightarrow~~~
q^{i}{}_{\underline{i}} = \delta^i_{\underline{i}} \re^{-U}
~~~\Longleftrightarrow~~~
q_{i}{}^{\underline{i}} = -\delta_i^{\underline{i}} \re^{-U}
~~~\Longleftrightarrow~~~
q_{i\underline{i}} = \ve_{i\underline{i}} \re^{-U}
~,
\label{SU2-gauge-fixing}
\ee
\esubeq
which breaks SU(2)$_R$.
After imposing the previous gauge fixing conditions, the remaining fundamental fields in the multiplet match 
those of the hyper-dilaton Poincar\'e supergravity multiplet \cite{Muller_hyper:1986ts} as summarised in table \ref{dof3}.
\begin{table}[hbt!]
\begin{center}
\begin{tabular}{ |c c c c c c c c c c |} 
 \hline
${e_m}^a$ & ${\omega_m}{}^{cd}$ & $A_{m}$ & $(\psi_m{}^\a_i,\bar{\psi}_m{}_\ad^i)$ & $W_{ab}$ 
& $(\rho_{\alpha}^{\underline{i}},\bar{\rho}^{\ad}_{\underline{i}})$ & $U$ & $b_{mn}{}^{\underline{i}\underline{j}}$ & $X^{ij}$ & $v_m$\\ 
$16B$ & $0$ & $4B$ & $32F$ & $6B$ & $8F$ & $1B$ & $18B$ & $3B$ & $4B$\\
\hline
$P_a$ & $M_{ab}$  & 
& $Q$ & {} & ${}$ & ${}$ & 
($\lambda_m{}^{\underline{i}\underline{j}}$) & {} & 
($\l_V$)\\
$-4B$ & $-6B$ & 
& $-8F$ & {} & {} & {} & $-9B$ & {} & $-1B$\\
\hline
\multicolumn{10}{|c|}{Result: $32+32$ degrees of freedom}\\
\hline
\end{tabular}
\caption{\footnotesize{Hyper-Dilaton Poincar\'e multiplet. Row $1$ gives all fields in the multiplet. Row two gives the number of independent components of these fields. Row three gives the surviving gauge symmetries. Row four gives the number of gauge degrees of freedom to be subtracted when counting the total degrees of freedom. Row five gives the resulting degrees of freedom. The parameter $\lambda_m{}^{\underline{i}\underline{j}}$ describes the vector symmetry associated with the 
triplet of gauge two-form $b_{mn}{}^{\underline{i}\underline{j}}$. The gauge parameter $\l_V$ describes the scalar symmetry of $v_m$.} \label{dof3}}
\end{center}
\end{table}
The fundamental fields are the vielbein ${e_m}^a$,
the gravitini $(\psi_m{}^\a_i,\bar{\psi}_m{}_\ad^i)$,
a real vector field $A_{m}$,
a real antisymmetric tensor $W_{ab}$,
a real scalar field that plays the role of a dilaton $U$,
a real triplet of scalar fields $X^{ij}$,
a triplet of gauge two forms $b_{mn}{}^{\underline{i}\underline{j}}$,
a gauge field $v_m$ that plays the role of the graviphoton,
and spinor fields $(\rho_{\alpha}^{\underline{i}},\bar{\rho}^{\ad}_{\underline{i}})$.
The residual gauge transformations of the multiplet 
are described by covariant general coordinate transformations
($\xi^a$),
local Lorentz transformations ($\l_{ab}$),
local supersymmetry ($\xi^\a_i,\bar{\xi}_\ad^i$),
and Abelian scalar $(\l_V)$ and vector $(\l_m{}^{\underline{i}\underline{j}})$ gauge transformations.
Note that we have kept the distinction of SU(2)$_R$ and SU(2) flavour indices. 
However, thanks to the second gauge condition in 
\eqref{SU2-gauge-fixing}, 
after gauge fixing the two indices can be identified.

The transformation rules of the resulting Poincar\'e supergravity multiplet 
\cite{Muller_hyper:1986ts}
are those that preserve the previous gauge conditions \eqref{gauge-conditions}.
To preserve the gauge condition
\eqref{fixing-phi} we need to impose 
$\l_{\mathbb D}\equiv0$ and $\l_{Y}\equiv0$.
Since $Q$-supersymmetry do not preserve the gauge, it is necessary 
to accompany these transformations with appropriate $S$-supersymmetry, special conformal, and SU(2)$_R$
compensating transformations.
To preserve the gauge condition \eqref{S-susy-gauge}, by examining the transformations \eqref{transf-lambda}
and \eqref{transf-lambda-bar}, it is straightforward to show that any $Q$-supersymmetry transformation
has to be accompanied by a compensating  $S$-supersymmetry transformation with parameter
\bsubeq
\bea
\eta_\alpha^i (\xi)
&=& 
-\hf(\s^{cd}\xi^{i})_\a\Big(F^+_{cd}+\hf W^+_{cd}\Big)
+\frac{1}{8}\xi_{\alpha j}X^{ji} 
+(\sigma^a\bar{\xi}^i)_\a A_a
~,
\\
\bar{\eta}^\ad_{i}(\xi)
&=&
~~\,
\hf(\ts^{cd}\bar{\xi}_i)^\ad \Big( F^-_{cd}+\hf W^-_{cd}\Big)
+\frac{1}{8}\bar{\xi}^{\dot{\alpha} j} X_{ji} 
-(\ts^a\xi_i)^\ad A_a
~.
\eea
\esubeq
A similar analysis shows that to preserve the gauge condition $b_m=0$ one needs to enforce nontrivial compensating 
special conformal $K$-transformations with a parameter $\l^a(\xi)$. However, since all the other supergravity fields
are conformal (not necessarily superconformal) primaries, not transforming under special conformal boosts, 
in practice we will never have to worry about inserting the compensating $\l^a(\xi)$ parameter (whose expression is quite involved)
in any Poincar\'e supergravity transformations.
The last gauge fixing condition which is not preserved is 
\eqref{SU2-gauge-fixing}.
It is straightforward to check that we can consistently have $\d q^{(i\underline{i})}=0$ by implementing in \eqref{d-qii} a 
compensating SU(2)$_R$ transformation with the following parameter
\bea
\lambda^{ij}(\xi)
= 
-\,\re^U\Big[
\xi^{(i}\rho^{j)} 
- \bar{\xi}^{(i}\bar{\rho}^{j)}
\Big]
~,
\eea
where 
$\rho^{i}=\d^i_{\underline{i}}\rho^{\underline{i}}$
and 
$\bar\rho_{i}=\d_i^{\underline{i}}\bar\rho_{\underline{i}}$.

At this stage, one has all the ingredients to obtain the 
transformation rules of any matter multiplet in the gauge fixed,
Poincar\'e supergravity frame,
by appropriately implementing the previous compensating 
gauge parameters in the superconformal transformation rules.
The resulting local transformation rules 
of the hyper-dilaton Poincar\'e multiplet form an algebra that
closes off-shell on a local extension of the $\cN=2$
super-Poincar\'e algebra with no residual $R$-symmetry. 
The structure of the algebra coincides, up to notation, 
with results in \cite{Muller_hyper:1986ts}.
A detailed presentation of the hyper-dilaton Poincar\'e multiplet 
and its coupling to matter in the superconformal
framework of this section will be given elsewhere as it is not
necessary for the rest of our paper.
It is worth underlining that, as explained by M\"uller 
in \cite{Muller_hyper:1986ts}, the resulting $32+32$ 
multiplet describes an irreducible off-shell representation.
This differs from
the case of the standard $40+40$ multiplet
of off-shell $\cN=2$ Poincar\'e supergravity 
\cite{Fradkin-Vasiliev-N2,deWit:1979xpv,deWit:1979dzm,Breitenlohner-Sohnius-N2}
which, for example,
can arise by coupling the standard Weyl multiplet to 
two compensators 
given by an off-shell vector and a hypermultiplet (the 
simplest of which is probably an off-shell linear multiplet).

\subsection{Hyper-Dilaton Poincar\'e supergravity action}
\label{chiral analysis}

With the hyper-dilaton Poincar\'e's multiplet 
recovered using a superconformal approach, 
we can now describe how to obtain the Poincar\'e supergravity action 
first constructed in \cite{Muller_hyper:1986ts}.
Once more, the construction is straightforward. 
In fact, the action derives from the kinetic action of the vector multiplet compensator in a hyper-dilaton Weyl multiplet
background after imposing 
the gauge fixing conditions \eqref{gauge-conditions}.
We will describe this construction by focusing only on the 
bosonic fields.

As a starting point we consider the bosonic sector of the chiral density formula for a system of vector multiplets possessing scalar fields $\phi^I$ with prepotential $\cF(\phi^I)$ in a standard Weyl multiplet background
\cite{deWit:1984rvr,deWit:1984wbb,Cremmer:1984hj,Butter_reduction:2012xg}.
This supersymmetric invariant has the following bosonic Lagrangian
\bea
e^{-1}\,\cL|_{bosonic} 
&=&
\mathcal{F}_I \square \bar{\phi}^I 
 + 3\mathcal{F}_I\bar{\phi}^I D 
 + \frac{1}{32}\mathcal{F}_{IJ} X^{I ij}X^J_{ij} 
 -\mathcal{F}_{IJ}f^{+I}{}^{ab}f^{+J}_{ab}
 -\hf\mathcal{F}\,W^-{}^{ab}W^-_{ab}
 \nonumber  \\
&&
-\mathcal{F}_IW^-{}^{ab}f^{-I}_{ab}
- \mathcal{F}_{IJ}\bar{\phi}^IW^{+}{}^{ab}f^{+J}_{ab} 
-\frac{1}{4}\mathcal{F}_{IJ}\bar{\phi}^I\bar{\phi}^J W^{+}{}^{ab}W^{+}_{ab} 
+{\rm c.c.}
~,
\label{chiral-bosonic}
\eea
where
$\cF_I=\frac{\pa\cF(\phi)}{\pa\phi^I}$ and $\cF_{IJ}=\frac{\pa^2\cF(\phi)}{\pa\phi^I\pa\phi^J}$.
We refer the reader to equation $(3.30)$ of \cite{Butter_reduction:2012xg} for a derivation of the previous Lagrangian.
In our case, we have only one vector multiplet and the only possible function $\cF$ we can choose which leads to a locally 
superconformal invariant is given by 
\be
\mathcal{F}(\phi) = -\frac{1}{4}\phi^2
~.
\label{function-compensator}
\ee
Here the overall factor is chosen for later convenience.
Once we insert \eqref{function-compensator} into \eqref{chiral-bosonic} and take into consideration
that we are working with a hyper-dilaton Weyl multiplet rather than a standard Weyl multiplet (meaning that 
\eqref{D_def} has to be used), we obtain the following 
\bea
e^{-1}\,\cL|_{bosonic} 
&=& 
-\frac{1}{4} |\phi|^2R
- \frac{|\phi|^2}{q^2}
q_{i\underline{i}}\cD^a\cD_a q^{i\underline{i}} 
-\frac{1}{2}\phi\cD^a\cD_a \bar{\phi} 
- \frac{1}{64} X^{ij}X_{ij}
 +\hf f^{+}{}^{ab}f^{+}_{ab}
\non\\
&&
+\hf\phi W^-{}^{ab}f^{-}_{ab}
+\hf\bar{\phi}W^{+}{}^{ab}f^{+}_{ab} 
+\frac{1}{8}\phi^2 W^-{}^{ab}W^-_{ab}
+\frac{1}{8}\bar{\phi}^2W^{+}{}^{ab}W^{+}_{ab} 
+{\rm c.c.}
~.~~~~~~~~
\eea
Note that in the previous Lagrangian there is 
a dependence upon 
the triplet of gauge two-forms 
$b_{mn}{}^{\underline{i}\underline{j}}$
which is still hidden in the SU(2)$_R$ connection inside the 
$\cD_a$ derivatives, see eq.~\eqref{compositeSU2}.

The final step to obtain the bosonic sector of the Poincar\'e supergravity of \cite{Muller_hyper:1986ts}
is to impose the gauge fixing conditions
\eqref{gauge-conditions}.
Upon implementing these conditions, the resulting Poincar\'{e} 
supergravity Lagrangian
turns out to be
\bea
e^{-1}\,\cL|_{bosonic}  
&=& 
 -\frac{1}{2}R
 + \frac{1}{2}f^{ab}f_{ab}
 + W^{ab}f_{ab}
 +\frac{1}{4}W^{ab}W_{ab}
-2(\partial_m U) \partial^m U 
 \nonumber \\
&&
+ 16\, \re^{4U} \tilde{h}^a{}_{kl} \tilde{h}_a{}^{kl} 
- \frac{1}{32} X^{ij}X_{ij}
+ 4A^aA_a 
~.
\label{hyper-Muller-action}
\eea
Note that here
$b_{mn}{}^{kl}
=
\d^k_{\underline{k}}\d^l_{\underline{l}}
b_{mn}{}^{\underline{k}\underline{l}}$ 
and
$\tilde{h}_a{}^{kl}
=
\d^k_{\underline{k}}\d^l_{\underline{l}}
\tilde{h}_a{}^{\underline{k}\underline{l}}$ 
since we have stopped distinguishing between underlined and 
non-underlined SU(2) indices after gauge fixing.

The structure of our and M\"{u}ller's Lagrangian 
in \cite{Muller_hyper:1986ts} coincide up to change of notation.
It is a straightforward
exercise to derive the fermionic extension of the previous
Lagrangian. This result will be presented elsewhere
together with a discussion of more general supergravity-matter 
couplings based on the hyper-dilaton Weyl multiplet
and the associated hyper-dilaton Poincar\'e supergravity.

To conclude let us analyse the on-shell structure of 
\eqref{hyper-Muller-action}.
It is clear that $W_{ab}$, $X^{ij}$, and $A_a$ are auxiliary 
fields that can be algebraically integrated out by using the 
equations of motion
\bea
W_{ab}=-2f_{ab}
~,~~~
X^{ij}=0
~,~~~
A_a=0
~.
\eea
Once the previous equations are used 
in \eqref{hyper-Muller-action},
one obtains the on-shell Lagrangian
\bea
e^{-1}\,\cL|_{bosonic}  
&=& 
 -\frac{1}{2}R
 - \frac{1}{2}f^{mn}f_{mn}
-2(\partial_m U) \partial^m U 
+ 16\, \re^{4U} \tilde{h}^m{}_{kl} \tilde{h}_m{}^{kl} 
~.
\label{hyper-Muller-action-on-shell}
\eea
The first two terms describe the standard kinetic terms
for minimal on-shell $\cN=2$ Poincar\'e supergravity
with a dynamical graviton and graviphoton.
The last two terms describe a dilaton and a triplet of dynamical
gauge two-forms which are not part of the minimal
on-shell $\cN=2$ Poincar\'e supergravity multiplet. In fact,
these fields describe the bosonic 
sector of an on-shell hypermultiplet where three of the scalars
have been dualised into real gauge two-forms 
\cite{Muller_hyper:1986ts}. The same holds by including the 
fermionic sector.

\subsection{$BF$-coupling and dilaton potential}
\label{section-3.3}

To conclude this section we consider an extension of 
the original result of M\"uller from \cite{Muller_hyper:1986ts}
and show how to construct by using superconformal techniques
a new off-shell supersymmetric invariant
that, e.\,g.,
leads to a non-trivial scalar potential for the dilaton.

Given a vector multiplet and a linear multiplet,
we consider the local supersymmetric
extension of a $BF$-action in 
a standard Weyl multiplet background \cite{deWit:1980gt}.
We refer the reader to \cite{Butter_reduction:2012xg}
for a derivation of the locally superconformal invariant, 
including fermionic terms,
in the notation used in our paper.
The bosonic part of such an invariant is
\bsubeq\label{BF}
\bea
e^{-1}\cL_{BF}|_{bosonic}  
&=&
F \phi 
+\bar{F} \bar{\phi}
+ \frac{1}{4} G_{i j} X^{i j} 
- 2\ve^{m n p q} b_{m n} f_{p q} 
~,
\\
&=&F \phi 
+\bar{F} \bar{\phi}
+ \frac{1}{4} G_{i j} X^{i j} 
-8\tilde{h}^{m} v_m
~.
\eea
\esubeq
By construction, the supersymmetric $BF$-action is 
also well defined as an invariant in a 
hyper-dilaton Weyl background.
We can readily construct an invariant of this form by 
considering the off-shell vector multiplet compensator
used in this section
and an off-shell linear multiplet given by
\bsubeq\label{composite-linear-xi}
\bea
&G_\xi{}_{ij}
:=
\xi_{\underline{i}\underline{j}}\,
G_{ij}{}^{\underline{i}\underline{j}}
~,~~~
\chi_\xi{}_{\alpha i}
:=\xi_{\underline{i}\underline{j}}\,
\chi_{\alpha i}{}^{\underline{i}\underline{j}}
~,~~~
\bar\chi_\xi{}^{\ad i}
=\xi^{\underline{i}\underline{j}}\,
\bar\chi^{\ad i}{}_{\underline{i}\underline{j}}
~,
\\
&
F_\xi
:=
\xi_{\underline{i}\underline{j}}\,
F^{\underline{i}\underline{j}}
~,~~~
\bar F_\xi=\xi^{\underline{i}\underline{j}}\,
\bar F_{\underline{i}\underline{j}}
~,~~~
b_\xi{}_{mn}
:=\xi_{\underline{i}\underline{j}}\,
b_{mn}{}^{\underline{i}\underline{j}}
~,~~~
\tilde{H}_\xi{}^a
=\xi_{\underline{i}\underline{j}}\,
\tilde{H}^a{}^{\underline{i}\underline{j}}
~.
\eea
\esubeq
Here 
$G_{ij}{}^{\underline{i}\underline{j}}$,
$\chi_{\alpha i}{}^{\underline{i}\underline{j}}$,
$\bar\chi^{\ad i}{}_{\underline{i}\underline{j}}$,
$F^{\underline{i}\underline{j}}$,
$\bar F_{\underline{i}\underline{j}}$,
$b_{mn}{}^{\underline{i}\underline{j}}$, 
and
$\tilde{H}^a{}^{\underline{i}\underline{j}}$
are  fields of the composite triplet of linear multiplets
\eqref{composite-linear} constructed in terms of 
fundamental fields of the hyper-dilaton Weyl multiplet,
while 
\be
\xi_{\underline{i}\underline{j}}
=
\xi_{\underline{j}\underline{i}}
~,~~~
(\xi_{\underline{i}\underline{j}})^*
=
\xi^{\underline{i}\underline{j}}
~,
\ee 
is a real triplet of
(structure group invariant) 
constants.
The bosonic part of the resulting Lagrangian is
\bea
e^{-1}\cL_{\xi}|_{bosonic}  
=
\xi_{\underline{i}\underline{j}}
\Big(
\frac{1}{4}
q_{i}{}^{\underline{i}}q_{j}{}^{\underline{j}}X^{i j} 
- 2\ve^{m n p q} b_{m n}{}^{\underline{i}\underline{j}} f_{p q} 
\Big)
=
\xi_{\underline{i}\underline{j}}
\Big(
\frac{1}{4}
q_{i}{}^{\underline{i}}q_{j}{}^{\underline{j}}X^{i j} 
-8\tilde{h}^{m}{}^{\underline{i}\underline{j}} v_m
\Big)
~.~~~~~~~~~
\label{Lxi}
\eea
After imposing the gauge fixing conditions
\eqref{gauge-conditions},
and adding the previous term into
\eqref{hyper-Muller-action},
we obtain the following Lagrangian
\bea
e^{-1}\,\cL|_{bosonic}  
&=& 
 -\frac{1}{2}R
 + \frac{1}{2}f^{ab}f_{ab}
 + W^{ab}f_{ab}
 +\frac{1}{4}W^{ab}W_{ab}
-2(\partial_m U) \partial^m U 
+ 4A^aA_a 
 \nonumber \\
&&
+ 16\, \re^{4U} \tilde{h}^a{}_{kl} \tilde{h}_a{}^{kl} 
- 2\xi_{ij}\,\ve^{m n p q}b_{m n}{}^{ij} f_{p q} 
- \frac{1}{32} X^{ij}X_{ij}
+\frac{1}{4}\xi_{ij}\,\re^{-2U}X^{i j} 
~,~~~~~~
\label{deformed-L}
\eea
where, after gauge fixing, we have used
$\xi_{ij}=\d_i^{\underline{i}}\d_j^{\underline{j}}
\xi_{\underline{i}\underline{j}}$
and $b_{m n}{}^{ij}
=
\d^i_{\underline{i}}\d^j_{\underline{j}}
b_{m n}{}^{\underline{i}\underline{j}}$.
As for the undeformed Lagrangian \eqref{hyper-Muller-action},
$W_{ab}$, $X^{ij}$, and $A_a$ are auxiliary 
fields that can be algebraically integrated out.
With the $\xi$-deformation turned on,
the equations of motion obtained from \eqref{deformed-L} are
\bea
W_{ab}=-2f_{ab}
~,~~~
X^{ij}= - 4\, \xi^{i j} \re^{-2U}
~,~~~
A_a=0
~.
\eea
Once these equations are used 
in \eqref{deformed-L},
we obtain the on-shell Lagrangian
\bea
e^{-1}\,\cL|_{bosonic}  
&=& 
 -\frac{1}{2}R
 - \frac{1}{2}f^{mn}f_{mn}
-2(\partial_m U) \partial^m U 
+ 16\, \re^{4U} \tilde{h}^m{}_{kl} \tilde{h}_m{}^{kl} \nonumber 
\\&&+ \,\xi^2\,\re^{-4U} 
+ 2 \xi_{ij}\, \ve^{mnpq} b_{mn}{}^{ij} f_{pq}
~,
\label{hyper-Muller+xi}
\eea
where
\be
\xi^2:=\hf \xi^{ij}\xi_{ij}
\geq0
~.
\ee
The first line coincides with the on-shell hyper-dilaton Poincar\'e 
supergravity \eqref{hyper-Muller-action-on-shell}
containing the standard minimal on-shell 
$\cN=2$ Poincar\'e supergravity
coupled to a dilaton and a triplet of dynamical
real gauge two-forms. 
Interestingly, 
the $\xi$-deformation induces a scalar potential for the
dilaton together with
a $BF$-coupling between the graviphoton and one of
the three gauge two-forms of the hyper-dilaton Poincar\'e multiplet 
(the component 
$b_\xi{}_{mn}=\xi_{\underline{i}\underline{j}}\,
b_{mn}{}^{\underline{i}\underline{j}}$
parallel to the $\xi_{ij}$ direction).
We now conclude by commenting the results obtained in this
subsection.

By considering a flat limit with 
$e_m{}^a\to\d_m^a$, 
$b_{mn}{}^{\underline{i}\underline{j}}\to0$, $U\to 0$, and
$q_i{}^{\underline{i}}\to\d_i^{\underline{i}}$,
and by keeping dynamical the vector multiplet with auxiliary
field $X^{ij}$, the Lagrangian \eqref{Lxi} turns into
\bea
\cL^{\rm flat}_{\xi}|_{bosonic}  
=
\frac{1}{4}
\xi_{ij}X^{i j} 
~.
\eea
This is a standard (electric) 
Fayet--Iliopoulos (FI) term 
for an $\cN=2$ vector multiplet
\cite{Fayet:1974jb,Fayet:1975yi}.
The invariant \eqref{Lxi} can be considered as 
a curved extension of a FI term in a hyper-dilaton Weyl
multiplet background. 
If one were to choose a different
gauge fixing to Poincar\'e supergravity where
$q^i{}_{\underline{i}}=\d^i_{\underline{i}}$
(a condition that would lead to the same model 
but in a string frame),
the dependence upon the dilaton would disappear from the 
term linear in $X^{ij}$.
This straightforwardly shows that if one restricts to 
a sector with constant dilaton,
the $\xi$-deformation leads to a negative cosmological 
constant $\L=-\xi^2\leq0$. Hence, the deformed model 
\eqref{hyper-Muller+xi} admits an ${\rm AdS}_4$ vacuum
with constant negative curvature proportional to $\xi^2$.

Another interesting aspect to comment about
is how the ${\rm SU(2)}_R$ symmetry 
plays a sharply different role for the
FI terms in off-shell 
$\cN=2$ supergravity based on the standard Weyl multiplet
compared to the 
hyper-dilaton Weyl multiplet case 
and the $\xi$-deformed M\"uller
supergravity described above.
When working with the standard Weyl multiplet 
(and even vector-dilaton Weyl multiplets), 
there is a close interplay between the ${\rm SU(2)}_R$
symmetry, the gauging of isometries of scalar field 
manifolds,
and the emergence of non-trivial scalar potentials.
Within the superconformal tensor calculus, 
this was already noticed in early investigations
of systems of Abelian vector multiplets \cite{deWit:1984rvr,Cremmer:1984hj},
and then extended to general  hypermultiplet sigma-models
\cite{deWit:2001brd,deWit:2001bk}.\footnote{See also  \cite{DAuria:1990qxt,Andrianopoli:1996vr,Andrianopoli:1996cm,DallAgata:2003sjo,Trigiante:2016mnt} for 
discussions concerning gauging and scalar potentials
by using alternative on-shell supergravity approaches.} 
By working with the standard Weyl multiplet, 
the ${\rm SU(2)}_R$ connection is an auxiliary field.
In the presence of FI terms, its equations of motion
identify the $R$-symmetries 
of the theory with the 
symmetries gauged by the $\cN=2$ vector multiplets.
This leads to non-trivial scalar potentials together with 
charges and masses for the gravitini ---
see \cite{VanProeyen:2004xt,Lauria:2020rhc} for reviews. 
The simplest case is the one of the standard Weyl multiplet
coupled to a single vector multiplet 
and a single hypermultiplet compensator.
In this case, the scalar potential is a simple negative cosmological constant and an FI term identifies on-shell
the $R$-symmetry connection with the graviphoton.
In  contrast, by using the hyper-dilaton Weyl multiplet,
the ${\rm SU(2)}_R$ connection is a composite field.
After gauge fixing, 
on-shell the $R$-symmetry is completely broken, its connection
is identified with the field strength of a dynamical
gauge two-form and the $\xi$-deformation introduces 
a dynamical $BF$-coupling.
As a result, one has 
an alternative procedure to obtain 
non-trivial scalar potentials compared to a setting
based on gaugings in an $\cN=2$ 
standard Weyl multiplet.\footnote{If one considers
higher-derivative interactions, 
it is possible to construct very general 
scalar potentials without gauged
$R$-symmetry by using a standard Weyl multiplet
and new types of $\cN=2$ FI terms, see 
\cite{Antoniadis:2019hbu}.}
It will be worth exploring this 
mechanism for more general off-shell 
matter systems coupled to a hyper-dilaton Weyl multiplet,
potentially including more physical hypermultiplets.

\section{Conclusion and future directions}
\label{section-4}

In our paper 
we have defined a new $24+24$ so-called
hyper-dilaton Weyl multiplet of $\mathcal{N} = 2$
conformal supergravity in four dimensions. 
The construction is based on reinterpreting the equations of motion
for an on-shell hypermultiplet 
as constraints that render some of the
fields of the standard Weyl multiplet composite.
By coupling the hyper-dilaton Weyl multiplet
to an off-shell vector multiplet compensator,
we have obtained a minimal $32+32$ off-shell
multiplet of $\cN=2$ Poincar\'e supergravity
that was constructed by M\"uller in \cite{Muller_hyper:1986ts}
and then, by using superconformal techniques, 
we have shown how to 
reproduce the supergravity action of \cite{Muller_hyper:1986ts}.
This contains the minimal on-shell $\cN=2$ Poincar\'e supergravity
coupled to a hypermultiplet where one
of its physical scalars plays the role of a dilaton 
while its three other scalars are dualised 
to a triplet of real gauge two-forms.
We have then described how a superconformal 
$BF$-coupling induces 
a scalar potential for the dilaton without a standard gauging.
There are several future directions that 
our work is opening up.
In the following we are going to mention a few.

As mentioned in the introduction, 
vector-dilaton Weyl multiplets 
were used in the past to study off-shell supergravity in five and six
space-time dimensions, see 
\cite{Bergshoeff:2001hc,Bergshoeff:1985mz} 
for descriptions in terms of component fields
and also \cite{Kuzenko:2008wr,Butter:2014xxa,BKNT16,Butter:2018wss} 
for analyses in superspace.
We are currently working towards extending our construction for  
hyper-dilaton Weyl multiplets in other ${\rm D}\leq6$ space-time dimensions.
Note also that much of the results obtained in our paper
were obtained by using the conformal superspace approach 
to $\cN=2$ conformal supergravity described in 
\cite{Butter_superspace:2011sr,Butter_reduction:2012xg}.
We will present superspace analyses together with more detailed
derivations of our results, and extensions to ${\rm D}\leq6$ dimensions
in the near future.

One of the main motivations of our work was to explore
alternative, yet simple, 
off-shell engineering of non-trivial scalar
potentials in 4D, $\cN=2$ supergravity. 
The results in subsection \ref{section-3.3}
are a first step in this direction.
While we have only presented in this paper an off-shell Poincar\'e
supergravity based on the hyper-dilaton Weyl multiplet
coupled to a single off-shell vector multiplet compensator,
a straightforward generalisation, 
part of a current work in progress, 
is to look at generic systems of (Abelian) vector multiplets.
It is well known that for these systems,
non-trivial scalar potentials in 4D, $\cN=2$
supergravity are associated to Fayet-Iliopoulos (FI) terms. 
These couplings are known to take two forms, 
either electric or magnetic FI terms.
The electric and magnetic nomenclature arise 
from the role that 
extensions of electro-magnetic duality of Maxwell theory
play in 4D, $\cN=2$ supersymmetry.
In the case of global supersymmetry, electric and 
magnetic FI terms are well understood both on-shell and off-shell,
see
\cite{Antoniadis:1995vb,IZ1,IZ2,RT,Antoniadis:2017jsk,Antoniadis:2019gbd,Kuzenko:2017gsc}.
They play an important role in the description of 
spontaneous full and partial breaking of supersymmetry.
They are also key ingredients in supergravity descriptions of 
compactified string theories with fluxes and various patterns of 
supersymmetry breaking, see e.\,g., \cite{Louis:2012ux} and references 
therein.
 In supergravity, the off-shell description 
of 4D, $\cN=2$ magnetic FI terms (and magnetic gaugings) 
has not been developed in full generality yet, 
though they are expected to play an important role in engineering 
scalar potentials in supergravity models 
possessing vacua with both positive and negative 
cosmological constant -- see for instance the
recent discussion of magnetic 4D, $\cN=1$ FI terms 
\cite{Antoniadis:2020qoj}.
The curved superspace constraints 
for off-shell magnetic FI terms were 
introduced in \cite{Kuzenko:2013gva,Kuzenko:2015rfx}
and in depth supergravity analyses in components 
(though not fully off-shell) 
were presented earlier in \cite{deVroome:2007unr,deWit:2011gk}.
By using a hyper-dilaton Weyl multiplet it is straightforward
to engineer generic electric and magnetic FI-type terms
by means of composite linear multiplets. 
We have already described how supergravity extensions 
of electric $\xi$-deformations can be obtained by using 
the $BF$-coupling \eqref{BF} in terms of the 
composite linear multiplet \eqref{composite-linear-xi}.
Off-shell magnetic FI-type deformations 
in a hyper-dilaton Weyl multiplet background
can easily be engineered in terms 
of the same composite linear multiplet. 
This would, for example, appear as an imaginary deformation of
the $X^{ij}$-auxiliary real
field of a vector multiplet.
Such deformations would be parametrised by the composite field
$G_\z{}_{ij}=\zeta_{\underline{i}\underline{j}}q_i{}^{\underline{i}}q_j{}^{\underline{j}}$ with 
$\zeta_{\underline{i}\underline{j}}=\zeta_{\underline{j}\underline{i}}$, $(\zeta_{\underline{i}\underline{j}})^*=\zeta^{\underline{i}\underline{j}}$
constants that generalise the magnetic FI terms of global 
supersymmetry.
Given a system of $n+1$ vector multiplets 
with scalar fields $\phi^I$ (with $I=0,1,\cdots,n$)
coupled to the off-shell 
hyper-dilaton Weyl multiplet, 
it is then straightforward to introduce $3(n+1)$ 
off-shell deformations each associated to either
a $\xi_I^{\underline{i}\underline{j}}$ electric deformation
or a $\z^I_{\underline{i}\underline{j}}$ magnetic deformation.
These induce non-trivial scalar potentials and vacuum structures.
We plan to report in the near future on work in progress 
based on this direction and to extend these analyses also 
by including more physical hypermultiplets.

Up until now, dilaton Weyl multiplets for ${\rm 4D}$ $\cN=2$ conformal supergravity have been constructed by coupling the standard Weyl multiplet to either an on-shell vector multiplet \cite{Butter:2017pbp} or an on-shell hypermultiplet (the latter in our current paper). It is quite clear that other variant dilaton Weyl multiplets might exist. A natural possibility is to couple the standard Weyl multiplet to either an on-shell linear (tensor) multiplet  or an on-shell vector-tensor multiplet -- see, e.g., \cite{SSW-2,SSW,deWKLL,Claus,HOW,DKT,Kuzenko:2011md,Butter_reduction:2012xg} for references on the vector-tensor multiplet including its coupling to conformal supergravity. It would be interesting to make these constructions explicitly and explore the peculiarities of these possible other dilaton Weyl multiplets in the study of off-shell ${\rm 4D}$, $\cN=2$ Poincar\'e supergravity.

Another natural direction for future research
is the construction of 
 higher-derivative actions based on the hyper-dilaton Weyl 
 and the hyper-dilaton Poincar\'e multiplets.
 Higher-derivative supergravity naturally arise in the low-energy
 description of string theory but, despite its importance, it is
 still poorly understood. 
Vector-dilaton Weyl multiplets 
have been successfully used to construct several
off-shell higher-derivative supergravities in $4\leq{\rm D}\leq 6$
dimensions, see
\cite{BSS1,CVanP,Bergshoeff:2012ax,OP131,OP132,OzkanThesis,Butter:2014xxa,NOPT-M17,Butter:2018wss,Mishra:2020jlc}.
It is natural to look at this problem starting from a 
hyper-dilaton Weyl multiplet
coupled to systems of vector multiplets with electric and magnetic FI-type terms. 
We expect to be able to overcome some of the other Weyl multiplet's restrictions to engineer off-shell gauged supergravity.

\vspace{0.3cm}
\noindent
{\bf Acknowledgements:}\\
We are grateful to 
I. Antoniadis, 
D. Butter, 
J.-P. Derendinger, 
J. Hutomo,
H. Jiang, 
S. Kuzenko, 
A. Van Proeyen,
and J. Woods
for discussions related to this work.
This work is supported by the Australian Research Council (ARC)
Future Fellowship FT180100353, and by the Capacity Building Package of the University
of Queensland.
G.G. and S.K. are supported 
by the postgraduate scholarships 
at the University of Queensland.


\appendix

\section{Notation and Conventions}

Throughout the paper we follow the 4D notation and conventions used in \cite{Buchbinder-Kuzenko} and \cite{Butter_reduction:2012xg}. We
summarize them here and include a number of useful identities.

The Minkowski metric is $\eta_{ab} = \textrm{diag}(-1, 1,1,1)$ and the four-dimensional sigma matrices are
\be
(\s^a)_{\a\ad} = (1, \vec{\s}) \ , \quad (\tilde{\s}^a)^{\ad\a} = \ve^{\ad\bd} \ve^{\a\b} (\s^a)_{\b\bd} = (1 , - \vec{\s})
~.
\ee
They satisfy
\begin{align}
(\s_a)_{\a\bd} (\tilde{\s}_b)^{\bd\b} = - \eta_{ab} \d^\b_\a - 2 (\s_{ab})_\a{}^\b \ , \\
(\tilde{\s}_a)^{\ad\b} (\s_b)_{\b\bd} = - \eta_{ab} \d^\ad_\bd - 2 (\tilde{\s}_{ab})^\ad{}_\bd \ ,
\end{align}
together with the following useful identities
\begin{subequations}
\bea
&(\s^a)_{\a \ad} (\s_a)_{\b \bd} = - 2 \ve_{\a\b} \ve_{\ad \bd}~,~~~
(\s_{ab})_{\a\b} (\s^{ab})_{\g \d} = -2 \ve_{\g (\a} \ve_{\b) \d}~, ~~~
(\s_{ab})_{\a\b} (\tilde{\s}^{ab})_{\gd \dd} = 0~, ~~~~~~~~~\\
&\ve_{abcd} (\s^{cd})_{\a \b} = -2 \ri (\s_{ab})_{\a\b} \ , \qquad \ve_{abcd} (\tilde{\s}^{cd})_{\ad \bd} = 2 \ri (\tilde{\s}_{ab})_{\ad\bd} \ ,\\
&\tr(\s_{ab} \s_{cd}) = (\s_{ab})_\a{}^\b (\s_{cd})_\b{}^\a = - \eta_{a[c} \eta_{d] b} - \frac{\ri}{2} \ve_{abcd}~,
\\
&\tr(\tilde{\s}_{ab} \tilde{\s}_{cd}) = (\tilde{\s}_{ab})^\ad{}_\bd (\tilde{\s}_{cd})^\bd{}_\ad = - \eta_{a[c} \eta_{d] b} + \frac{\ri}{2} \ve_{abcd}
~.
\eea
\end{subequations}
Here the Levi-Civita tensor $\ve^{abcd}$ obeys
\be \ve^{0123} = - \ve_{0123} = 1 
~,~~~~~~
\ve^{abcd} \ve_{a' b' c' d'} = - 4! \d^a_{[a'} \d^b_{b'} \d^c_{c'} \d^d_{d']}~,
\ee
where (anti-)symmetrization of $n$ index includes a $1/n!$ normalization. For example, given a bi-vector  $A_{ab}$ and a bi-spinor $B_{\a\b}$ tensors,
it holds
\be A_{[ab]} = \frac{1}{2!} (A_{ab} - A_{ba}) \ , \quad B_{(\a\b)} = \frac{1}{2!} (B_{\a\b} + B_{\b\a}) \ .
\ee

For four-dimensional spinors we use the two-component notation.
Dotted and undotted spinor indices are raised and lowered using the following conventions
\be
\phi_\a = \ve_{\a\b} \phi^\b \ ,~~~
\phi^\a = \ve^{\a\b} \phi_\b \ ,~~~~~~
\quad \bar{\psi}^\ad = \ve^{\ad\bd} \bar{\psi}_\bd \ ,
\quad \bar{\psi}_\ad = \ve_{\ad\bd} \bar{\psi}^\bd \ ,
\ee
where the epsilon matrices obey
\be \ve_{\a\b} = - \ve_{\b\a} \ , \quad \ve_{\a\b} \ve^{\b\g} = \d_\a^\g \ , \quad \ve_{\ad\bd} \ve^{\bd\gd} = \d_\ad^\gd \ , \quad \ve^{12} =- \ve_{12} = 1 \ .
\label{epsilon-identities}
\ee
Similarly, we raise and lower ${\rm SU}(2)$ indices using the conventions
\be
\phi_{i} = \ve_{i j} \phi^j \ ,~~~
\phi^i = \ve^{i j} \phi_j\ ,
\ee
where $\ve^{ij}$ and $\ve_{ij}$ satisfy the same relations \eqref{epsilon-identities} as the spinor ones.
Spinor indices are contracted as follow
\begin{align}
\phi \psi &:= \phi^\a \psi_\a \ , \quad \bar{\phi} \bar{\psi} = \bar{\phi}_\ad \bar{\psi}^\ad~.
\end{align}
A few examples of such contractions used in the paper are
\begin{subequations}
\bea
&\bar{\xi}_k \tilde{\s}^a \de_a \Sigma^{k} &= \bar{\xi}_{\ad k} (\tilde{\s}^a)^{\ad \a} \de_a \Sigma^{k}_{\a}~,  \\
&{\psi_a}^{ i}\sigma^a \tilde{\s}^{cd} \bar{\rho}^{\underline{i}} &=  {\psi_a}^{\a i}(\sigma^a)_{\a \ad} (\tilde{\s}^{cd})^{\ad}{}_{\bd} \bar{\rho}^{\bd \underline{i}}~, \\
&(\psi_{a}{}^{i}\s^{cd})^\a &= \psi_{a}{}^{\b i}(\s^{cd})_\b{}^\a 
~.
\eea
\end{subequations}

A vector $T_a$ can be rewritten with spinor indices as
\be
T_{\a\bd} = (\s^a)_{\a\bd} T_a \ , \quad T_a = - \hf (\tilde{\s}_a)^{\bd\a} T_{\a\bd} \ .
\ee
A real antisymmetric tensor, $W_{ab} = - W_{ba}$ is converted to spinor indices as follow
\be
W_{\a\b} = \hf (\s^{ab})_{\a\b} W_{ab} \ , \quad \bar{W}_{\ad\bd} = - \hf (\tilde{\s}^{ab})_{\ad\bd} W_{ab} \ ,\quad
W_{ab} = (\s_{ab})^{\a\b} W_{\a\b} - (\tilde{\s}_{ab})_{\ad\bd} \bar{W}^{\ad\bd} \ .
\ee
If $W_{ab}$ is real, then
$W_{\a\b}=W_{\b\a}$ and $\bar{W}_{\ad\bd}=\bar{W}_{\bd\ad}$ are complex conjugates of each others.


\begin{footnotesize}

\end{footnotesize}

\end{document}